\documentclass[preprint, 10pt]{elsarticle}
\usepackage{booktabs,amsmath,amssymb,latexsym,graphicx,color,amsfonts,enumerate,url,siunitx,setspace}
\usepackage{caption,mathtools,rotating,upgreek,gensymb,epstopdf,afterpage,xcolor,wrapfig,subfig,subfloat,stackengine,sidecap}
\usepackage[colorlinks]{hyperref}
\usepackage{leftidx}
\usepackage{natbib}
\biboptions{sort&compress}
\usepackage{float}

\usepackage[paperwidth=8.5in, paperheight=11.0in, top=1.1in, bottom=1.2in, left=1.1in, right=1.1in]{geometry}

\usepackage{tikz}
\newcommand*\circled[1]{\tikz[baseline=(char.base)]{
            \node[shape=circle,draw,inner sep=2pt] (char) {#1};}}
            
\let\today\relax
\makeatletter
\def\ps@pprintTitle{%
    \let\@oddhead\@empty
    \let\@evenhead\@empty
    \def\@oddfoot{\footnotesize\itshape
         {} \hfill\today}%
    \let\@evenfoot\@oddfoot
    }
\makeatother

\begin{document}
\bibliographystyle{elsarticle-num-names}

\begin{frontmatter}

\title{A Continuum Macro-Model for Bistable Periodic Auxetic Surfaces}

\author{E.~Sansusthy Tardio}
\ead{esansust@CougarNet.UH.EDU}

\author{T.~Chen}
\ead{tianchen@Central.UH.EDU}

\author{Th.~Baxevanis\corref{bla}}
\ead{tbaxevanis@uh.edu}
\cortext[bla]{Corresponding author}
\address{Department of Mechanical \& Aerospace Engineering, University of Houston, 
Houston, TX 77204-4006, USA}

\begin{abstract}
A macro-constitutive model for the deformation response of periodic rotating bistable auxetic surfaces is developed. Focus is placed on isotropic surfaces made of bistable hexagonal cells composed of six triangular units with two stable equilibrium states. Adopting a variational formulation, the effective stress–strain response is derived from a free energy function expressed in terms of the invariants of the logarithmic strain. To address the mathematical ill-posedness and numerical artifacts—such as mesh sensitivity—arising from the double-well nature of the free energy, two regularization approaches are introduced: (i) a gradient-enhanced first invariant of the logarithmic strain, and (ii) an artificial material rate dependency. Although neither regularization guarantees solution uniqueness, the former mitigates mesh sensitivity, while the latter improves the convergence behavior of the nonlinear numerical scheme by promoting smooth temporal evolution of transition localization and enabling the system to overcome snap-backs induced by local non-proportional loading near transition fronts. The model is implemented using membrane/shell structural elements and plane stress continuum ones within the ABAQUS finite element suite. Numerical simulations demonstrate the efficacy of the proposed formulation and its implementation.
\end{abstract}


\end{frontmatter}

\section{Introduction}\label{Intro}
Mechanical metamaterials are designed matter whose distinctive mechanical properties are primarily governed by their unique structural design rather than their chemical composition. Unlike conventional materials that become thinner in cross-section when stretched, auxetic metamaterials exhibit lateral expansion, leading to enhanced mechanical performance characteristics~\cite{wang2014auxetic,sanami2014auxetic,mir2014review,carneiro2013auxetic,yang2004review,ren2018auxetic,lim2015auxetic,scarpa2008auxetic,jiang20183d,bettini2010composite,jayanty2011auxetic}. Large-scale shape transformations in auxetic materials can be programmed by leveraging the complex deformation modes associated with elastic instabilities, requiring the pre-stressed state to be maintained in the structure, or, in order to by-pass the latter requirement, by designing metamaterials capable of continuously transitioning between stable states as needed. Bistability refers to the existence of two stable equilibrium states within a material or structure, allowing for reversible transitions between these states under external stimuli. Bistability arises from a nonconvex, multi-welled potential energy landscape that may give rise to mechanical hysteresis, which may absorb significant amounts of energy, and to interesting dynamic phenomena involving large-amplitude nonlinear wave motion~\cite{chi2022bistable,cao2021bistable,zhang2024review,scarselli2016novel,pellegrini2013bistable,kochmann2017exploiting}.  

Bistable auxetic materials open avenues for innovative applications across diverse fields, including aerospace, biomedical, robotic, and electronics industries. These materials offer a unique ability to manipulate and control mechanical wave propagation through the formation of band gaps—frequency ranges of strong wave attenuation—generated via Bragg scattering~\cite{hussein2014dynamics}. This capability opens the door to a wide range of applications, including wave guiding~\cite{khelif2004guiding, kafesaki2000frequency}, cloaking~\cite{cummer2007one}, and noise reduction~\cite{elser2006reduction, casadei2010periodic}. Additionally, large-amplitude nonlinear wave motion, driven by transitions between stable states~\cite{Betts2013, Yang2014}, can be harnessed for energy harvesting~\cite{Harne2013, Wu2014}, vibration control~\cite{Johnson2014}, and pulse propagation in lossy media~\cite{Nadkarni2016a, Nadkarni2016b}. Beyond these applications, bistable auxetics hold promise for biomedical advancements, such as controlled encapsulation and release of medicine~\cite{shim2012buckling}. Their ability to transition from a flat state to a desired target geometry also presents significant advantages in reducing fabrication, transport, and construction costs for morphing and deployable structures~\cite{chen2021bistable}.

\begin{SCfigure}[][t]
\centering
\includegraphics[width=0.7\columnwidth, keepaspectratio=true]{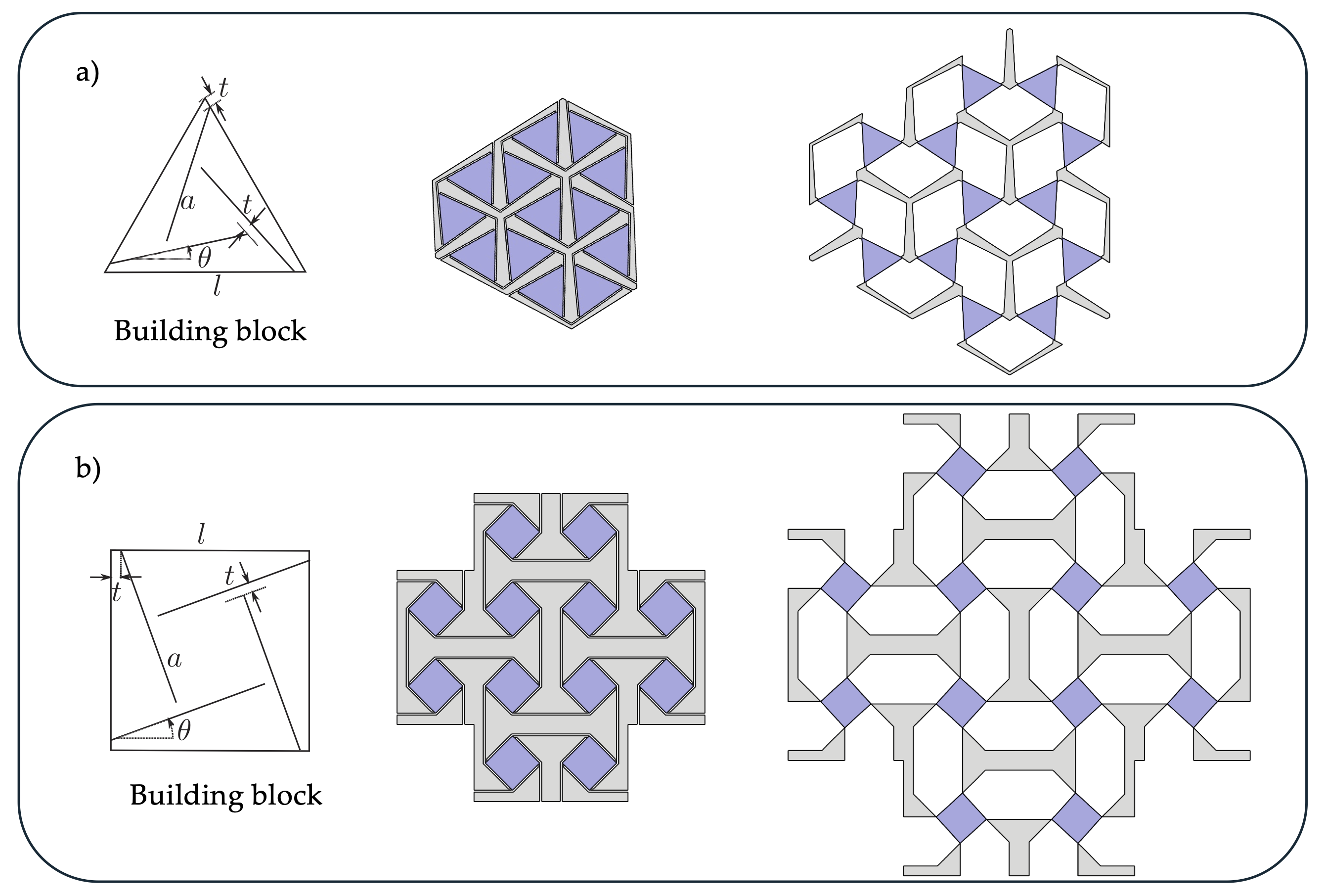}
\caption{Rotating bistable auxetics based on (a) equilateral triangular, and (b) square building blocks connected at their vertices via hinges along with the corresponding unit cells in their undeformed and stretched configuration~\cite{rafsanjani2016bistable}. }
\label{cells}
\end{SCfigure}

\begin{SCfigure}[][t]
\centering
\includegraphics[width=0.7\columnwidth, keepaspectratio=true]{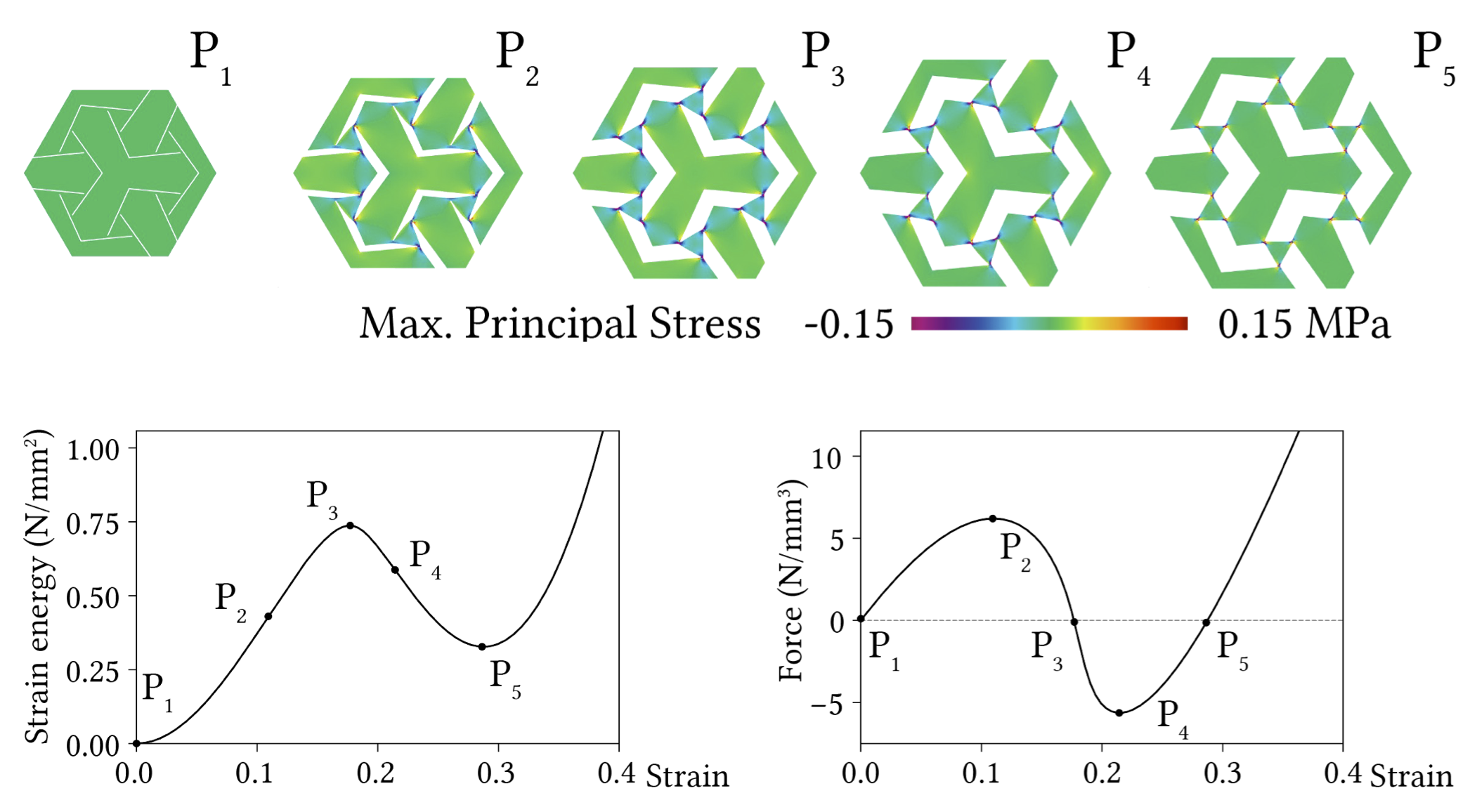}
\caption{The bistable hexagonal cell is composed of six triangular units. The stress concentration in the compliant hinge joints connecting the rotating inner triangle elements during stretching is depicted. The strain energy and force as a function of strain (subfigures, left and right, respectively, at the bottom) exhibit the typical bistable behavior. The two stable states correspond to points $P_1$ and $P_5$~\cite{chen2021bistable}.}
\label{response}
\end{SCfigure}

\begin{SCfigure}[][t]
\centering
\includegraphics[width=0.3\columnwidth, keepaspectratio=true]{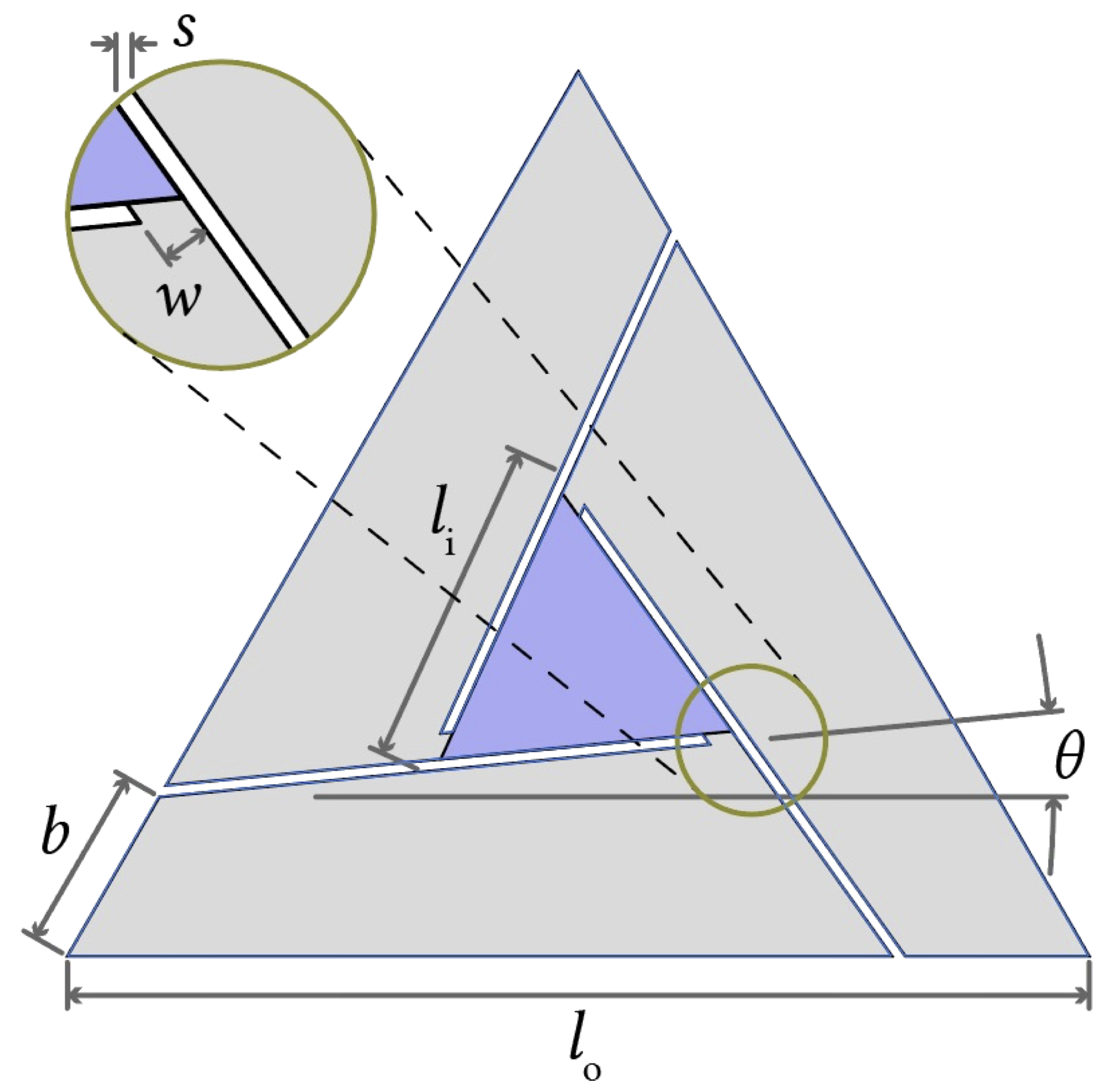}
\hspace{15pt}
\caption{The bistable hexagonal cell in Figure~\ref{response} is composed by six triangular units (a single triangle unit shown in this figure), each parameterized by a width $b$ and an angle $\theta$~\cite{chen2021bistable}.}
\label{triangle}
\end{SCfigure}

Recently, perforated rotating unit auxetics~\cite{grima2010perforated, shan2015design, grima2016auxetic} have been employed to create planar bistable auxetics~\cite{rafsanjani2016bistable, chen2021bistable}. These versatile structures are easily adaptable to various patterns and can be manufactured through simple cuts at different length scales, inspired by geometric motifs found in ancient architecture. These structures consist of equilateral triangles and square building blocks connected at their vertices via hinges (Figure~\ref{cells}). When stretched in one direction, the coordinated rotation of these units causes the material to expand transversely. Designs based on equilateral triangles allow for isotropic tuning of auxetic behavior with targeted expandability. Anisotropic rotating unit auxetics can be achieved by squares or rectangles. Their deformation response was examined both experimentally and via Finite Element Analysis (FEA) by pulling their opposite edges until full expansion was achieved. During this process, the specimens transitioned through metastable states and maintained their deformed shape even after the load was removed. The flexure hinges, seamlessly integrated into the structure, bent to facilitate relative rotation between adjacent units. This unique mechanism fused auxetic behavior with structural bistability. In contrast, conventional rotating unit auxetics return to their original configuration when the load is released. The behavior of a periodic unit cell under uniaxial tension, followed by release after achieving full expansion, is shown in Figure~\ref{response}. The simulation monitored the engineering stress–strain relationship, strain energy density, and Poisson’s ratio of the material. Initially, the stress–strain curve exhibited a linear response, but it transitioned into a strongly nonlinear regime, with the load dipping below zero before returning to positive values. This response indicates the presence of a local energy minimum at a non-zero deformation, corresponding to the material's second stable state. The Poisson’s ratio remained negative throughout the entire range of applied stretches and notably equaled -1 at the second stable state. An analytical approximation of the macro-stretch $\varepsilon_{b}$ required to reach the second stable state, based on geometric (pattern) parameters (Figure~\ref{triangle}), can be derived, which gives an insight into the effect of these pattern parameters on the bistable response of these materials~\cite{chen2021bistable}
\[
\varepsilon_{b} \approx \frac{2}{l_o} l_i \sin \left( \theta + \frac{\pi}{6} \right),
\]
where $l_o$, $l_i$, and $\theta$ are defined in Figure~\ref{triangle}.


Despite the potential of bistable auxetic materials, their fundamental mechanical behavior remains largely unexplored. The existing research addresses the basic performance of bistable auxetic materials, \emph{i.e.}, is limited to the proof-of-concept level, overlooking the critical interplay of material properties, structural geometry, and the external forces that influence and describe their response. This knowledge gap limits the ability to design bistable auxetic materials and structures that can reliably and predictably perform under diverse conditions, and restricts their broader application in industry and technology; therefore, the ability to control and optimize bistable auxetic responses is essential for creating advanced materials capable of precise and efficient mechanical performance and high resilience.

To date, the mechanical response of bistable periodic metamaterials has been modeled primarily through simplified spring–mass models or fully resolved FEA~\cite{Nadkarni2016b, Deng2020, Yasuda2020, jin2020guided, Bonthron2025, Paliovaios2024, Frazier2023, Zhang2022}. Although periodic rotating bistable auxetics can be modeled using purely kinematic approaches~\cite{friedrich2018locally, konakovic2016beyond, konakovic2018rapid}, incompatibilities between the unit triangles during stretching of the hexagonal cell indicate that localized strains develop in the hinges. Consequently, the bistable functionality depends critically on the elasticity of the underlying material (Figure~\ref{response}), necessitating detailed FEA studies. As these systems increase in size and architectural complexity, modeling them as fully resolved discrete structures becomes computationally prohibitive, particularly for design exploration and optimization. Effective macro-continuum formulations capable of efficiently predicting transition phenomena in bistable periodic metamaterials remain scarce. Recently, an effective continuum model was developed, specifically designed to simulate the dynamic propagation of stable transition waves triggered by snap-through between two equilibrium states~\cite{jin2020guided, khajehtourian2021continuum}. These waves are powered by the energy released as unit cells transition from the high-energy (open or deformed) configuration to the low-energy (closed or undeformed) equilibrium, causing the transition front to propagate in that direction. In the present work, an effective continuum model is developed that can capture both forward and reverse phase transitions, \emph{i.e.}, transitions from low- to high-energy equilibria and \emph{vice versa}, in periodic rotating bistable auxetic surfaces under displacement-controlled quasi-static loading. The proposed model operates efficiently at the continuum scale while incorporating the underlying structural architecture through a homogenized representation. The formulation is implemented using membrane, shell, and plane-stress continuum elements within the ABAQUS finite element suite.

\section{Macro-Effective (Homogenized) Constitutive Response}
A constitutive model for the effective deformation response of periodic rotating bistable auxetic surfaces is developed and implemented in finite elements assuming a hyperelastic base material. As in classical homogenization, a separation of scales between the unit cell size and the overall sample dimensions is assumed. 

The stress–strain response is derived from a proposed strain energy density potential, $\psi$, \emph{i.e.}, a free energy potential.  In general, $\psi$ for isotropic ``hyperelastic" materials can be defined by either the principal invariants $I_i$ $(i=1,2,3)$ of $\boldsymbol{C} (=\boldsymbol{F}^T\boldsymbol{F})$~\cite{rivlin1948large} or the principal stretches $\lambda_i$ of $\boldsymbol{V}$~\cite{ogden1972large, ogden1997non, ogden1986recent}, where $\boldsymbol{F}$ is the deformation gradient and $\boldsymbol{F}=\boldsymbol{V}\boldsymbol{R}$ stands for polar decomposition. For Ogden-type free-energy potentials of the form $\psi \equiv \psi(\lambda_1, \lambda_2, \lambda_3)$,  calibration through experimental data curve fitting is usually a challenge, and additionally the extrapolative predictions in loading scenarios not included in the calibration are poor and/or the response becomes unstable beyond the fitted range. Similar challenges persist for free-energy potential of the form $\psi \equiv \psi(I_1, I_2, I_3)$ since isolating the effect of each invariant for material parameter fitting purposes is not a straightforward task. Instead, a free-energy potential $\psi \equiv \psi(\bar{I}_1, \bar{I}_2, \bar{I}_3)$, where $\bar{I}_i$ stand for the invariants of the logarithmic strain tensor $\boldsymbol{h}\equiv \ln (\boldsymbol{V})$, as proposed by Criscione et al.~\cite{criscione2000invariant}, enable a more tractable path to introducing phenomenological fitting functions that capture experimental data since $\boldsymbol{h}$ has the advantage of additively separating dilatation from distortion and the derivatives $\partial \bar{I}_i/\partial \boldsymbol{h}$ are mutually orthogonal.

The rotating bistable auxetics are modeled as material surfaces. Such an idealization allows for (i) developing a concise and mathematically amenable theory for the in-plane deformation response, which is of primary interest for the majority of potential applications, and (ii) the development of structural elements, \emph{i.e.}, membrane and shell elements (under additional assumptions regarding the out-of-plane deformation response for the latter), and plane stress continuum ones that will, in turn, enable numerical simulations of interest. 

For material surfaces the following two invariants of the logarithmic strain are of interest
\begin{enumerate}
\item The first invariant $\bar{I}_1 = \text{tr}(\boldsymbol{h})= \ln J$ $(J=\det (\boldsymbol{F}))$, which represents the volume change of the material;
\item The second invariant $\bar{I}_2 = |\text{dev}(\boldsymbol{h})|=\sqrt{\left(\boldsymbol{h} - \frac{1}{2}\bar{I}_1\right)^2}$, which represents the magnitude of constant-volume material distortion.
\end{enumerate}

The in-plane Kirchhoff stress tensor is then given as
\begin{equation}\label{hypercl}
\boldsymbol{\tau} = \frac{\partial \psi}{\partial \boldsymbol{h}}=\frac{\partial \psi}{\partial \bar{I}_1}\boldsymbol{\delta} + \frac{\partial \psi}{\partial \bar{I}_2} \boldsymbol{N}  = \left(\frac{\partial \psi}{\partial \bar{I}_1} - \frac{1}{2}\frac{\partial \psi}{\partial \bar{I}_2}\frac{\bar{I}_1}{\bar{I}_2}\right)\boldsymbol{\delta} + \frac{1}{\bar{I}_2}\frac{\partial \psi}{\partial \bar{I}_2} \boldsymbol{h}, 
 \end{equation}
where for the free-energy potential $\psi$, the following assumption is adopted
\begin{equation}\label{psif}
\psi = c \bar{I}_2^2 + h(\bar{I}_1).
\end{equation}
Thus
\begin{equation}\label{psif1}
\frac{\partial \psi}{\partial \bar{I}_1} = h^\prime(\bar{I}_1) ,
\end{equation}
and
\begin{equation}\label{psif2}
\frac{\partial \psi}{\partial \bar{I}_2} = 2 c \bar{I}_2 ,
\end{equation}
where $c$ is a constant, prime denotes differentiation with respect to $\bar{I}_1$, the function $h$ is a polynomial of adequate order to fit the response of the auxetic bistable surfaces, and $\boldsymbol{\delta}$ is the unit tensor with components $\delta_{ij}=1$ if $i= j$ and $\delta_{ij}=0$ if $i \ne j$. Note that $ \partial \psi/\partial \bar{I}_1=p$, where $p$ stands for the hydrostatic Kirchhoff stress, and $\boldsymbol{N} $ is a deviatoric tensor.

\subsection{Gradient-enhanced and viscous regularization}

The double-well nature of the free-energy potential may result in loss of ellipticity of the governing equilibrium equations. When ellipticity is lost, discontinuous spatial derivatives may emerge along characteristic planes (\cite{zauderer2011partial}, pp. 135)–where the governing equations have turned hyperbolic–with significant consequences, \emph{e.g.}, finite element solutions can become unreliable and dependent on mesh size. One manifestation of the loss of ellipticity of the governing boundary value problems is that the size of transition fronts is arbitrarily narrow (\ref{SML}). Transitions fronts are diffuse boundaries, which separate topological domains comprised by open (strained) and closed (unstrained) unit cells. These fronts are formed by high-energy transitioning unit cells that collectively contribute to the interface energy. As external (quasi-static) loading changes, these domain boundaries shift, driving the propagation of transition fronts. To locally avoid loss of ellipticity and provide more robust and objective simulation results, various regularization methods have been proposed in the literature, such as non-local or gradient-enhanced continua~\cite{bazant1984continuum, 10.1115/1.3225725, peerlings1996some, fleck1993phenomenological, gurtin2003framework, forest2009micromorphic, shaat2020review}, and the addition of viscosity (material rate-dependence)~\cite{needleman1988material, loret1990dynamic, prevost1990dynamic, wang1996interaction, benallal2008note, 10.1115/1.2897246}. While material rate-dependence implicitly introduces a characteristic length scale into the governing boundary value problem for inelastic solids, this scale is dictated by the domain geometry, such as imperfections or inhomogeneities~\cite{needleman1988material, 10.1115/1.2897246}. In contrast, the gradient-enhanced regularization incorporates an intrinsic material length scale that governs the width of transition fronts. Nevertheless, such a formulation typically delays, but does not entirely prevents, the loss of ellipticity~\cite{aravas2021non, aravas2023implicit}. By comparison, material rate-dependence has been shown to eliminate mesh sensitivity in inelastic materials by ensuring solution uniqueness, not merely by smoothing strain discontinuities~\cite{needleman1988material, 10.1115/1.2897246, aravas2021non}. 

Here, the model’s response is regularized by introducing
\begin{description}

\item[a gradient-enhanced description] of the invariant, $\bar{I}_1$, in which the model’s response 
\begin{equation}\label{viscous_nl}
\boldsymbol{\tau} \overset{\eqref{psif1}, \eqref{psif2}}{:=} \left[h^\prime(\bar{I}_1^g) - c\bar{I}_1^g\right]\boldsymbol{\delta} + 2c\boldsymbol{h},
 \end{equation}
 is complemented by a modified Helmholtz equation
 \begin{equation}\label{NL}
\bar{I}_1=\bar{I}_1^g-l^2 \nabla\cdot \nabla \bar{I}_1^g, \ \ \left. \nabla \bar{I}_1^g \cdot \boldsymbol{n}_0\right|_{\partial \mathcal{B}_0}=0
\end{equation}
where the appended superscript $^g$ indicates that the invariant $\bar{I}_1^g$ is affected by gradient activity, $\mathcal{B}_0$ is the reference configuration of the continuum body of interest, $\boldsymbol{n}_0$ the unit normal to its boundary $\partial \mathcal{B}_0$, and $\nabla$ stands for the gradient operator. \eqref{NL} can be interpreted as a moving averaging operator acting on the trace of the logarithmic strain tensor $\bar{I}_1$ and delivering a smoothed field $\bar{I}_1^g$. 

\item[an artificial material rate-dependency] in the form
\begin{equation}\label{viscous}
\boldsymbol{\tau} = \left[h^\prime(\bar{I}_1^g) - c\bar{I}_1^g\right]\boldsymbol{\delta} + 2c\boldsymbol{h}+\eta \dot{\boldsymbol{h}},
 \end{equation}
where $\eta$ is a parameter with units of viscosity MPa $\cdot$ s, which should remain sufficiently small so that does not significantly alter the ``actual" rate-independent material response. 

\end{description} 

As discussed in Section~\ref{tension} and~\ref{SML}, the introduced artificial viscosity does not guarantee solution uniqueness or remove mesh dependence; nevertheless, it effectively complements the gradient regularization by enhancing the convergence behavior of the nonlinear numerical scheme. Rather than exhibiting an abrupt response jump, transition localization develops smoothly in time under the control of the viscosity parameter, which may further aid the algorithm in traversing snap-backs induced by locally non-proportional loading along transition fronts.

\subsection{Model calibration}\label{cal}

The material model is calibrated from the response of a unit cell, which is initially a hexahedron with flat, axis-aligned faces, under periodic boundary conditions assuming a hyperelastic base material (\ref{CRBM} and \ref{CPMM}). The fundamental macroscopic kinetic and kinematical measures in these simulations are the macroscopic 1st Piola–Kirchhoff stress tensor $\bar{\boldsymbol{P}}$ and the macroscopic deformation gradient $\bar{\boldsymbol{F}}$, respectively, defined in terms of the volume average of their microscopic counterparts, $\bar{\boldsymbol{P}}=\langle \boldsymbol{P} \rangle$, $\bar{\boldsymbol{F}}=\langle \boldsymbol{F} \rangle$, where $\langle \bullet \rangle = \frac{1}{\mathcal{V}_0}\int_{\mathcal{V}_0} \bullet \, dV$ stands for the volume average of the quantity $\bullet$ and $\mathcal{V}_0$ is the reference configuration of the unit cell. $\bar{\boldsymbol{P}}$ is determined based on the work balance criterion $ \bar{\boldsymbol{P}}:d\bar{\boldsymbol{F}}=\langle \boldsymbol{P} :d\boldsymbol{F} \rangle$~\cite{danielsson2002three} from displacement-driven simulations under periodic boundary conditions on the outer boundary of the unit cell $\partial \mathcal{V}_0$ 
\[
\boldsymbol{x}-\boldsymbol{x}'=\bar{\boldsymbol{F}}\left(\boldsymbol{X}-\boldsymbol{X}'\right) \ \ \text{(periodicity of deformation) on}\, \partial \mathcal{V}_0,\ \  \text{and}
\]
 \[
 \boldsymbol{P} \boldsymbol{n}_{0_V}=-\boldsymbol{P}' \boldsymbol{n}_{0_V}' \ \  \text{(anti-periodicity of tractions) on}\, \partial \mathcal{V}_0,
 \]
 where $\boldsymbol{x}$ and $\boldsymbol{X}$ are the position vectors for the current and reference configurations, respectively, $\boldsymbol{n}_{0_V}$ is the outward unit normal to a point $\boldsymbol{X} \in \partial \mathcal{V}_0$, and $'$ indicates the unique, periodically located on the boundary, points assigned to points on $\partial \mathcal{V}_0$ and quantities evaluated at those points. Specifically, macroscopic uniaxial loading
\[
\bar{\boldsymbol{F}}-\boldsymbol{\delta}=(\lambda-1)\boldsymbol{e}_1\otimes \boldsymbol{e}_1+H_{22}\boldsymbol{e}_2\otimes \boldsymbol{e}_2,
\]
is imposed on the unit cell for the calibration process, where $\lambda$ is a load parameter, $H_{22}$ is not prescribed but determined by the condition $\bar{P}_{22}=0$ and $\boldsymbol{e}_1$ and $\boldsymbol{e}_2$ are the unit vectors of the Cartesian coordinate system.

\subsection{Membrane/shell structural and plane-stress continuum elements in ABAQUS suite}\label{implem}

The in-plane membrane/shell/plane-stress response is governed by the equilibrium for a 3D body in a state of plane stress, \emph{i.e.}, the material surface model proposed with the additional assumption that the cross-section thickness of the elements remains unaltered irrespectively of the membrane strain in accordance with unit cell simulations; state transition is accommodated by strains developing in the hinges of the cells. For the shell elements, the transverse shear treatment is assumed linear elastic based on the initial elastic modulus of the base material and, thus, the transverse shear strain and force are assumed constant over the element. Therefore, all stiffness integration locations have the same transverse shear strain, transverse shear section force, and transverse shear stress distribution.  The transverse shear stiffness is specified as $K_{11}=5f_p \mu_0 t /6$, $K_{12}=0$, and $K_{22}=5 f_p \mu_0 t /6$, where $f_p=\left(1 + 0.25 \times 10^{-4}A/t^2\right)^{-1}$ is a dimensionless factor that is used to prevent the shear stiffness from becoming too large in thin shells, $A$ is the area of the element, $t$ is the shell thickness, $\mu_0$ is the material shear modulus, and the number 5/6 is the shear correction coefficient that results from matching the transverse shear energy to that for a three-dimensional structure in pure bending. The implementation of the Helmholtz-type equation~\eqref{NL} utilizes ABAQUS’s coupled temperature–displacement analysis, where the temperature field is reinterpreted as the non-local trace of the logarithmic strain tensor, $\bar{I}_1^g$. All constitutive equations are integrated via the UMAT subroutine interface.

\section{Numerical Simulations}

\subsection{Tension of a strip with a geometric imperfection}\label{tension}
\begin{figure}
\centering
\includegraphics[width=0.55\textwidth, keepaspectratio=true]{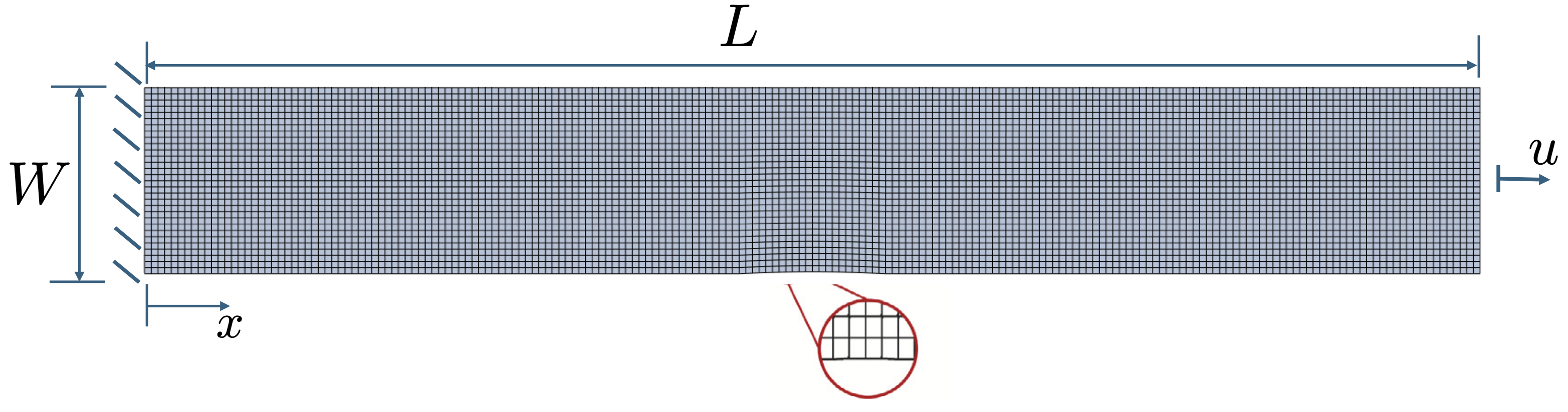}
\caption{Schematic of the strip geometry with the finite element mesh.}
\label{Strip}
\end{figure}

Tension simulations are performed on a strip with dimensions $L:W:l=10:1.5:0.15$ under plane stress conditions (Figure~\ref{Strip}). In the simulations, the axial displacement of all nodes on the left edge is zero and that of all the nodes on the right edge specified as $u$. The nodes at both the left and right edge are allowed to move freely in the transverse direction except at the bottom-left corner where both displacement degrees of freedom are constrained to prohibit rigid body motions. In order to initiate a localized state transition, a small geometric imperfection is introduced by means of a notch with 1\% reduced width, positioned in the middle of the bottom of the strip. The material parameters used in the calculations are listed in Table~\ref{tab:mpa}. These correspond to the unit cell geometry (UCG), with pattern parameters also listed in the same table, and the base material properties listed in Table~\ref{tab:mpbm}. Stress and strain are normalized in the presented results with respect to the maximum stress, $\sigma_t$, and the logarithmic strain, $h_t$, corresponding to $\sigma_t$ obtained just before softening in a uniaxial test (see Figure~\ref{Dots}).

\begin {figure}[]
\centering
\subfloat[Axial, volume-averaged, normalized stress–strain curve for the strip tension test for the mesh density of $L_e \sim  l/3$ using the gradient-enhanced and rate-dependent model.]{\label {strvsstr_a}\includegraphics[height=0.35\textwidth, keepaspectratio=true]{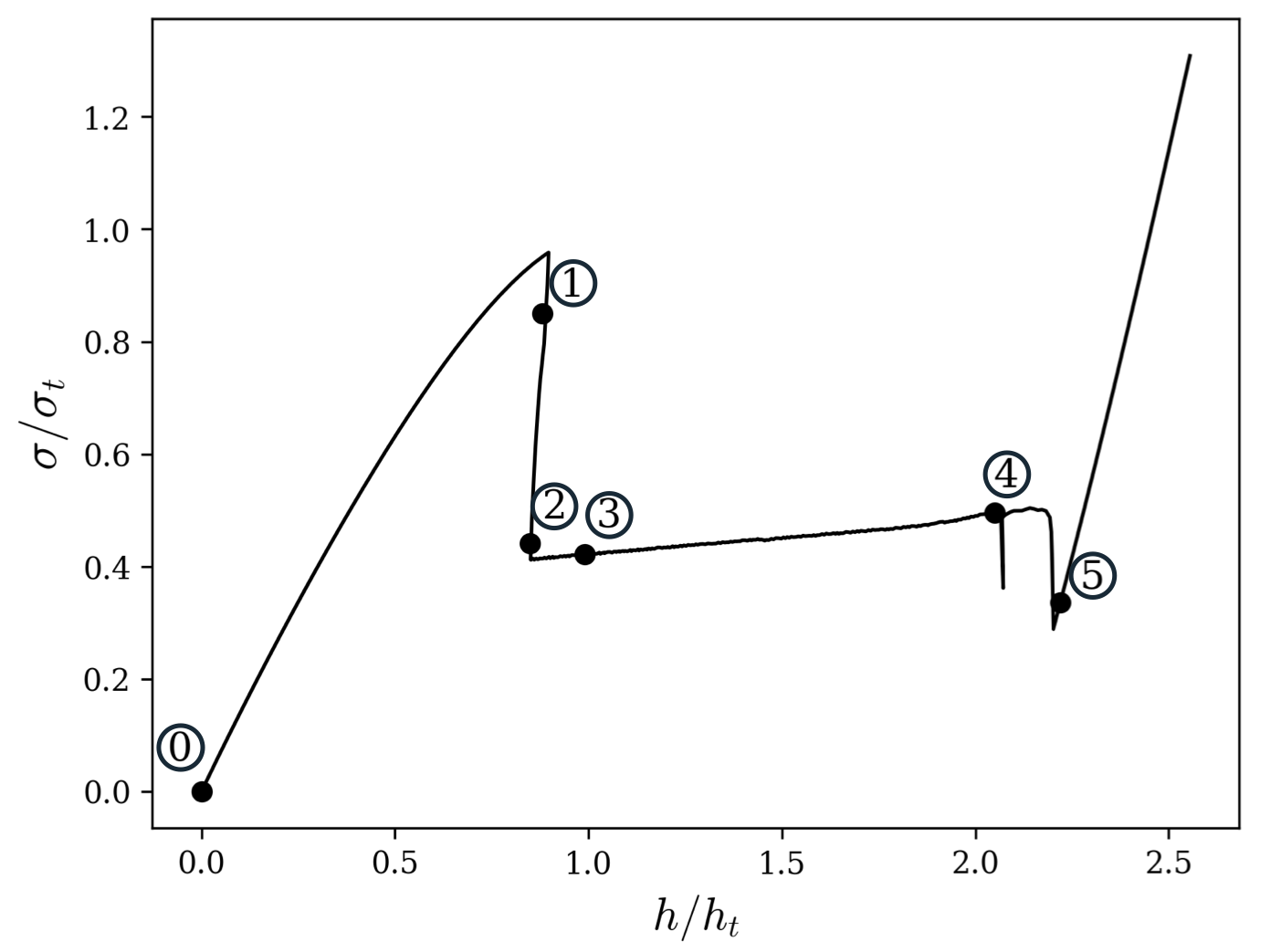}}\\
\subfloat[Controur plots of the normalized first invariant, $\bar{I}_1^g$ (left), and hydrostatic Kirchhoff stress, $p$ (right), in the deformed configuration of the strip at different loading instances.]{\label {strvsstr_b}\includegraphics[width=0.95\textwidth, keepaspectratio=true]{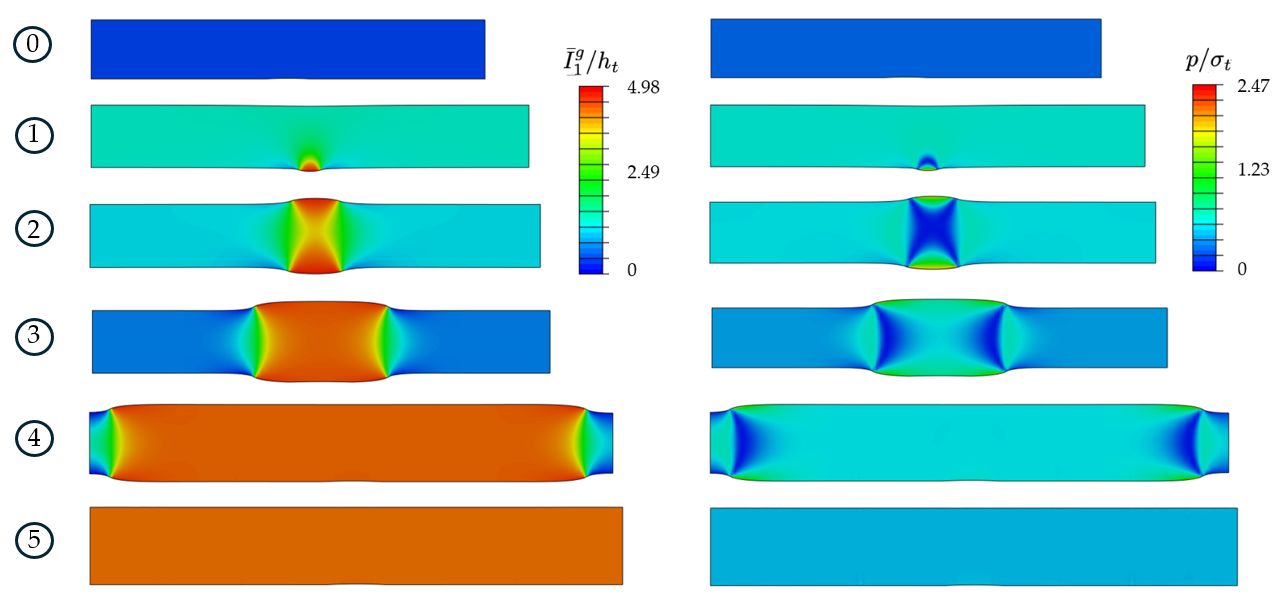}}
\quad
\vspace{0pt}
\caption{}
\label {contour}
\end {figure}

In the first simulation, the gradient-enhanced and rate-dependent model is used and the strip is discretized into 6,000 linear quadrilateral elements with an approximate element size of $L_e = l/3$, and the viscosity parameter is set to $\eta = 0.9$ MPa$\cdot$s. Figure~\ref{strvsstr_b} shows contour plots of the invariant, $\bar{I}_1^g$, and the hydrostatic Kirchhoff stress, $p$. Up to Point $\circled{2}$ in Figure~\ref{strvsstr_a}, the deformation remains largely uniform. Inhomogeneous deformation begins at the location of the geometric imperfection, triggered by local stress concentration at an axial strain of approximately $0.89h_t$, initiating a state transition. This onset of localized deformation is reflected in the nominal stress–strain response as a sharp drop in axial stress, marking the nucleation of transition bands. This drop continues until the transition front reaches the top edge of the strip. The slope of the drop is steeper than that corresponding to uniform uniaxial loading, as discussed in~\ref{SML}, with the material outside the band unloading. As the transition propagates horizontally toward the left and right edges, the nominal stress plateaus. A trailing depression zone follows the transition front, corresponding to the negative region of the material’s stress–strain curve, which must be traversed to complete the state transition. The two stress drops observed at the end of the plateau correspond to the instances when the transition fronts reach the lateral edges of the strip. Notably, despite the problem’s symmetry, the fronts do not advance simultaneously but alternate.

\begin {figure}[]
\centering
\subfloat[Gradient-enhanced and rate-dependent model.]{\label {strvs_a}\includegraphics[width=0.45\textwidth, keepaspectratio=true]{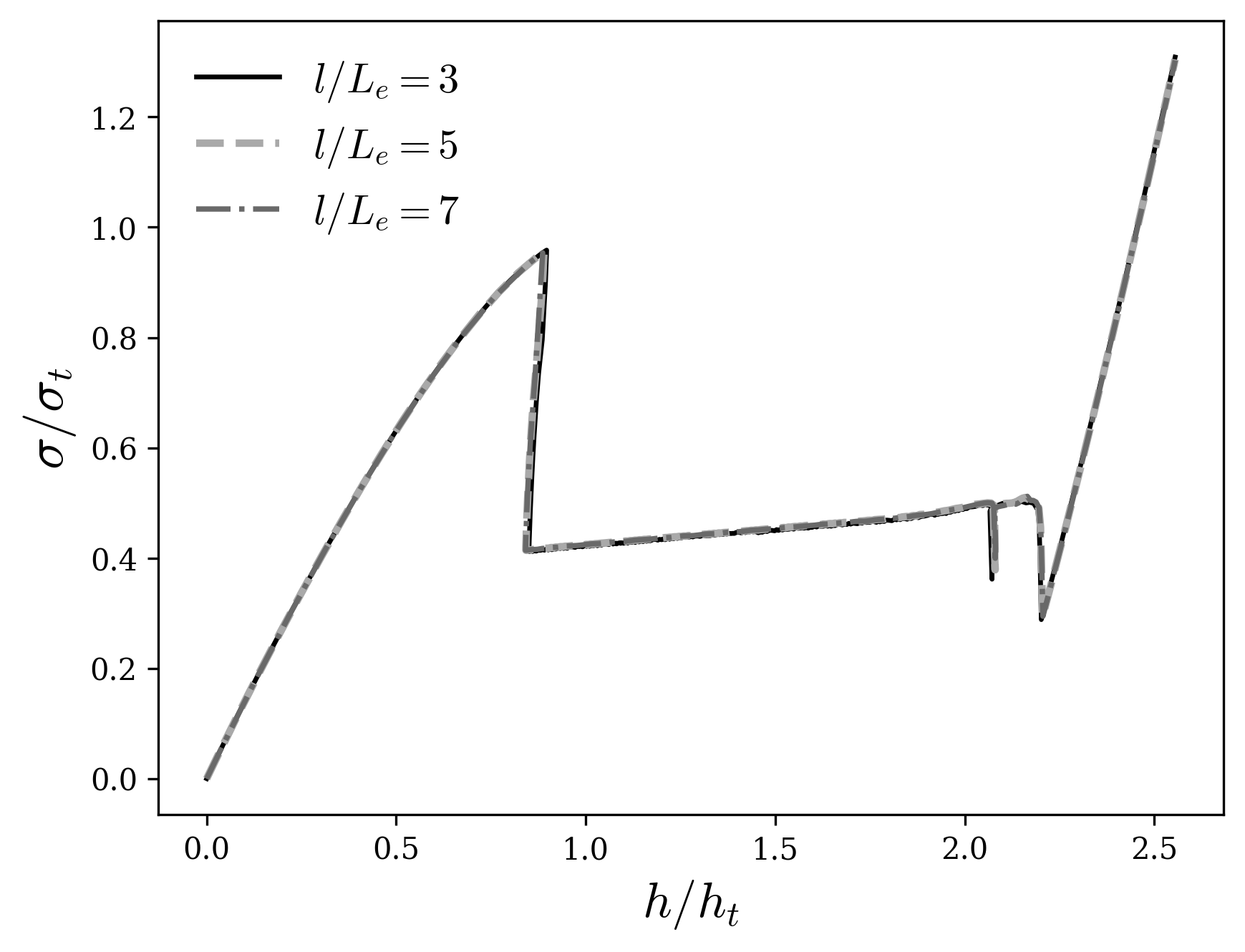}}
\quad
\subfloat[Enlarged views of the normalized $h_1$-distribution, where $h_1$ is the maximum principal logarithmic strain, in the band transition zone for mesh densities of $L_e=l/3$, $L_e=l/5$, and $L_e=l/7$ during front propagation.]{\label {h1band_a}\includegraphics[width=0.65\textwidth, keepaspectratio=true]{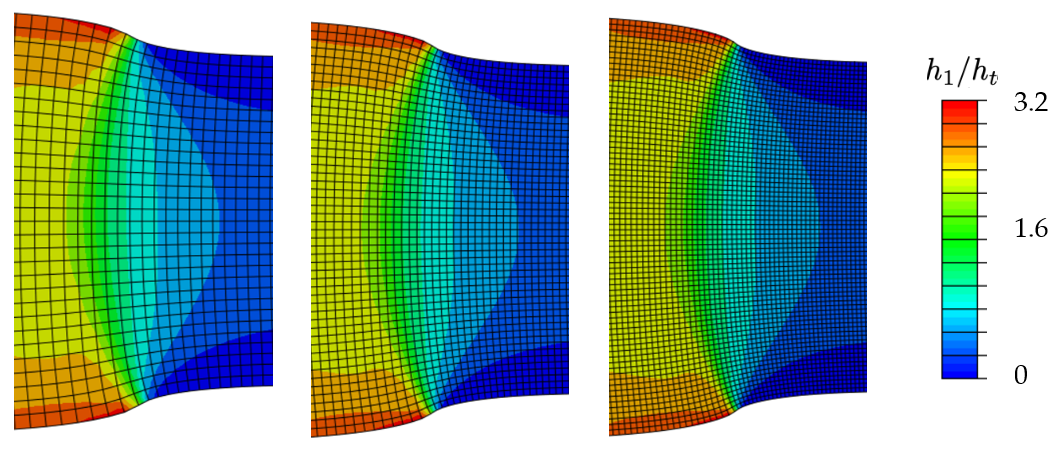}}
\quad
\subfloat[Normalized $h_1$  plotted along the centerline through the transition zone corresponding to the meshes shown in (b).]{\label {h1band_b}\includegraphics[height=0.35\textwidth, keepaspectratio=true]{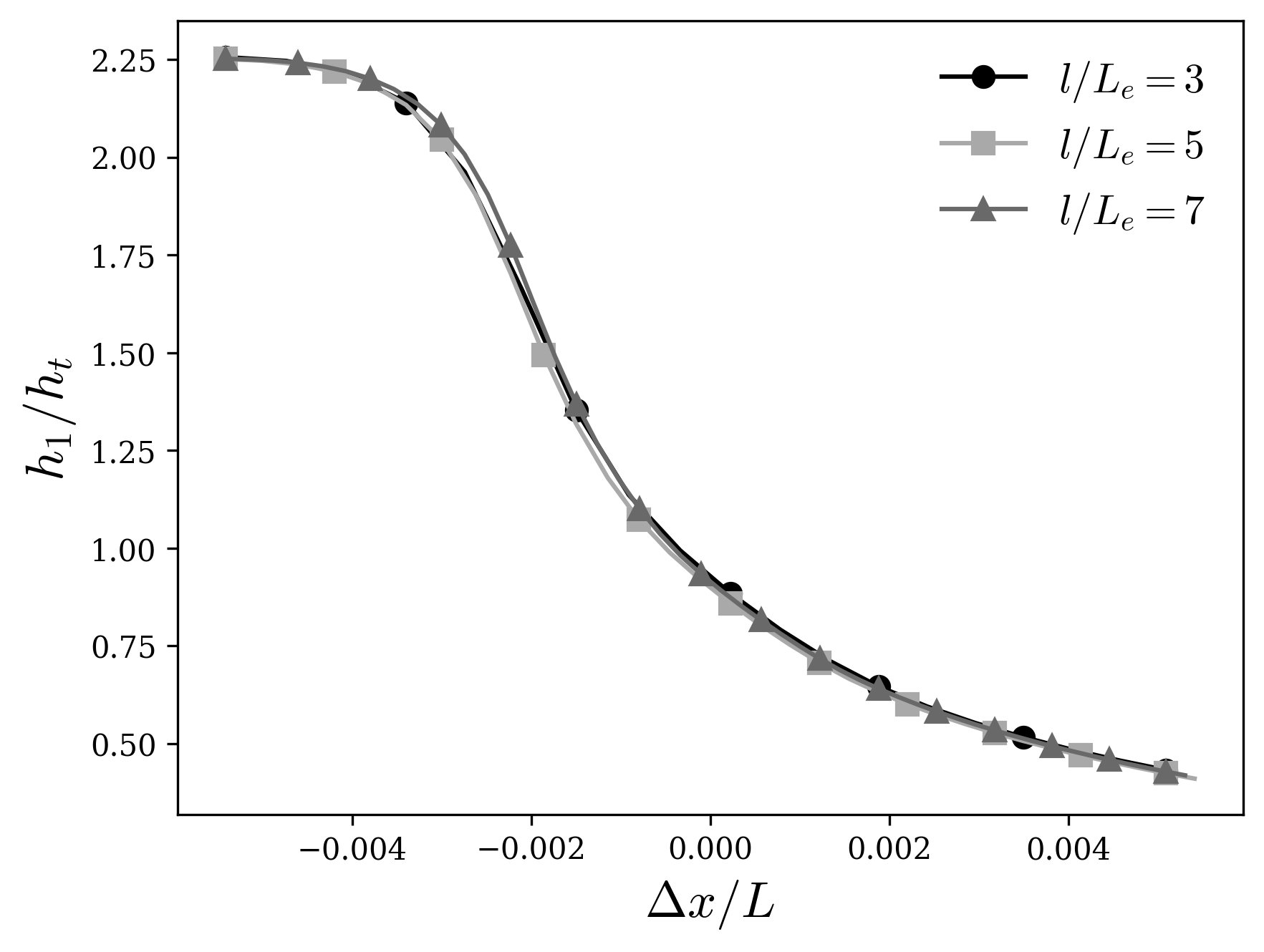}}
\caption{a) Axial, volume-averaged, normalized stress vs strain responses for the strip tension test for mesh densities of $L_e=l/3$, $L_e=l/5$, and $L_e=l/7$ for the gradient-enhanced and rate-dependent model;  b) \& c) band transition zone for the different mesh densities. }
\label {strvsstr}
\end {figure}

A mesh sensitivity analysis is presented next, along with a discussion of the influence of the adopted regularization on the simulation results. Axial stress–strain responses from two additional mesh densities are compared against the baseline response from the mesh used in the previous simulations (Figure~\ref{strvs_a}). The results show no significant deviation among the three curves, indicating mesh-independent global response. The mesh sizes and the corresponding viscosity parameters, $\eta$, used to ensure numerical convergence are listed in Table~\ref{tab:mesh}. To further verify mesh independence, snapshots of the strip with contours of the first invariant $\bar{I}_1^g$ are taken once the transition fronts have formed and propagated toward the lateral edges. In the magnified views, the transition zone spans multiple elements in each mesh, and both the width and shape of the transition front remain consistent across all densities (Figure~\ref{h1band_a}). This is corroborated by the $\bar{I}_1^g$–$x$ profile along the centerline, which shows a clearly defined transition zone extending over approximately 20, 33, and 46 elements for the respective meshes (Figure~\ref{h1band_b}). These findings confirm that the combined gradient and viscous regularization successfully mitigate mesh sensitivity. However, the mesh must be sufficiently fine to resolve strain gradients, particularly near the transition zone boundaries. The simulations suggest that an element size of $L_e \sim  l/3$ or smaller is adequate for consistent results.
\begin{figure}
\centering
\includegraphics[width=0.45\textwidth, keepaspectratio=true]{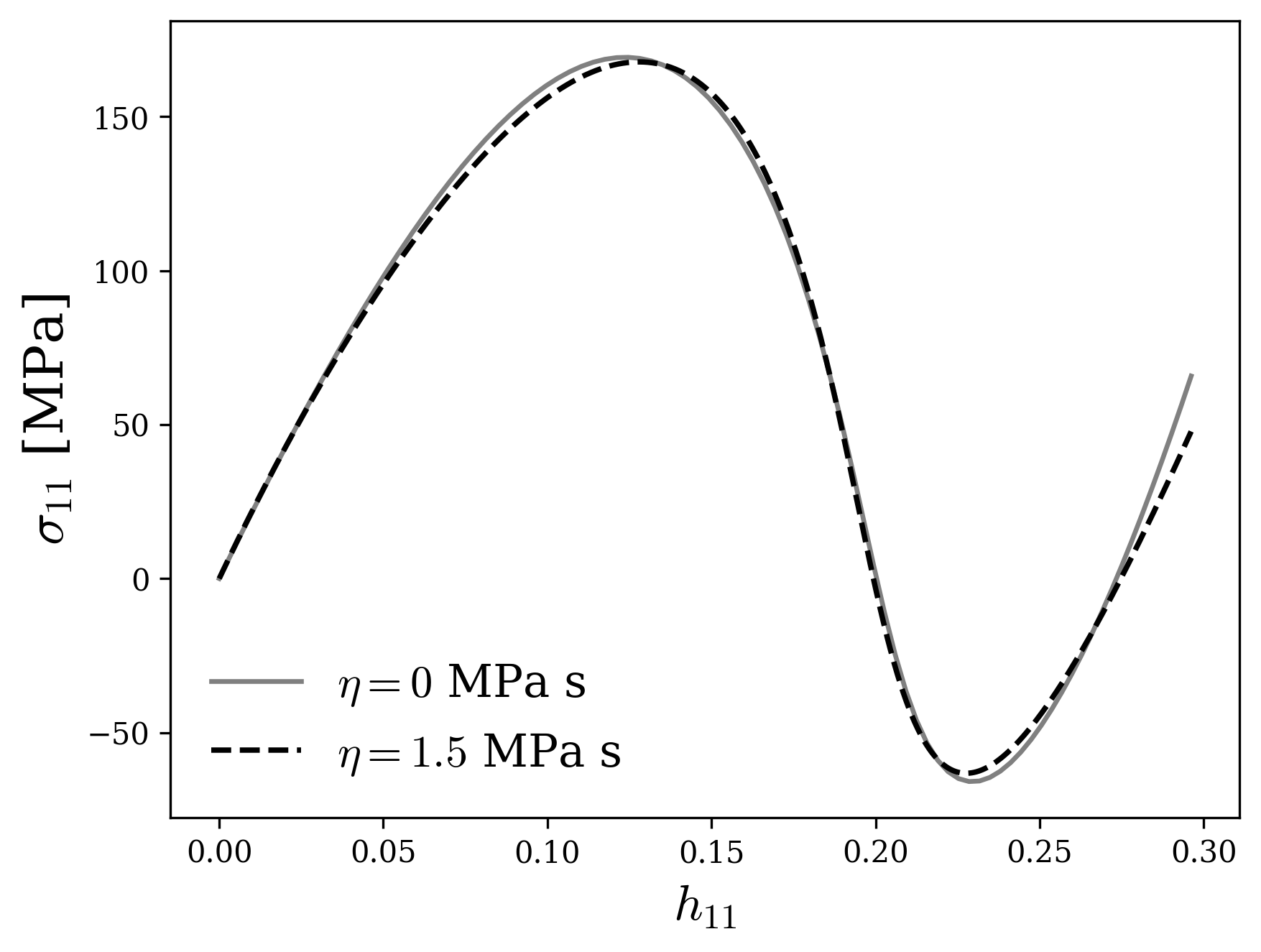}
\caption{Uniaxial stress vs strain response for the rate-independent model and the rate-dependent one with the maximum viscosity parameter, $\eta=1.5$ MPa . s, used in the numerical simulations.}
\label{eta}
\end{figure}
\begin{figure}
\centering
\includegraphics[width=0.45\textwidth, keepaspectratio=true]{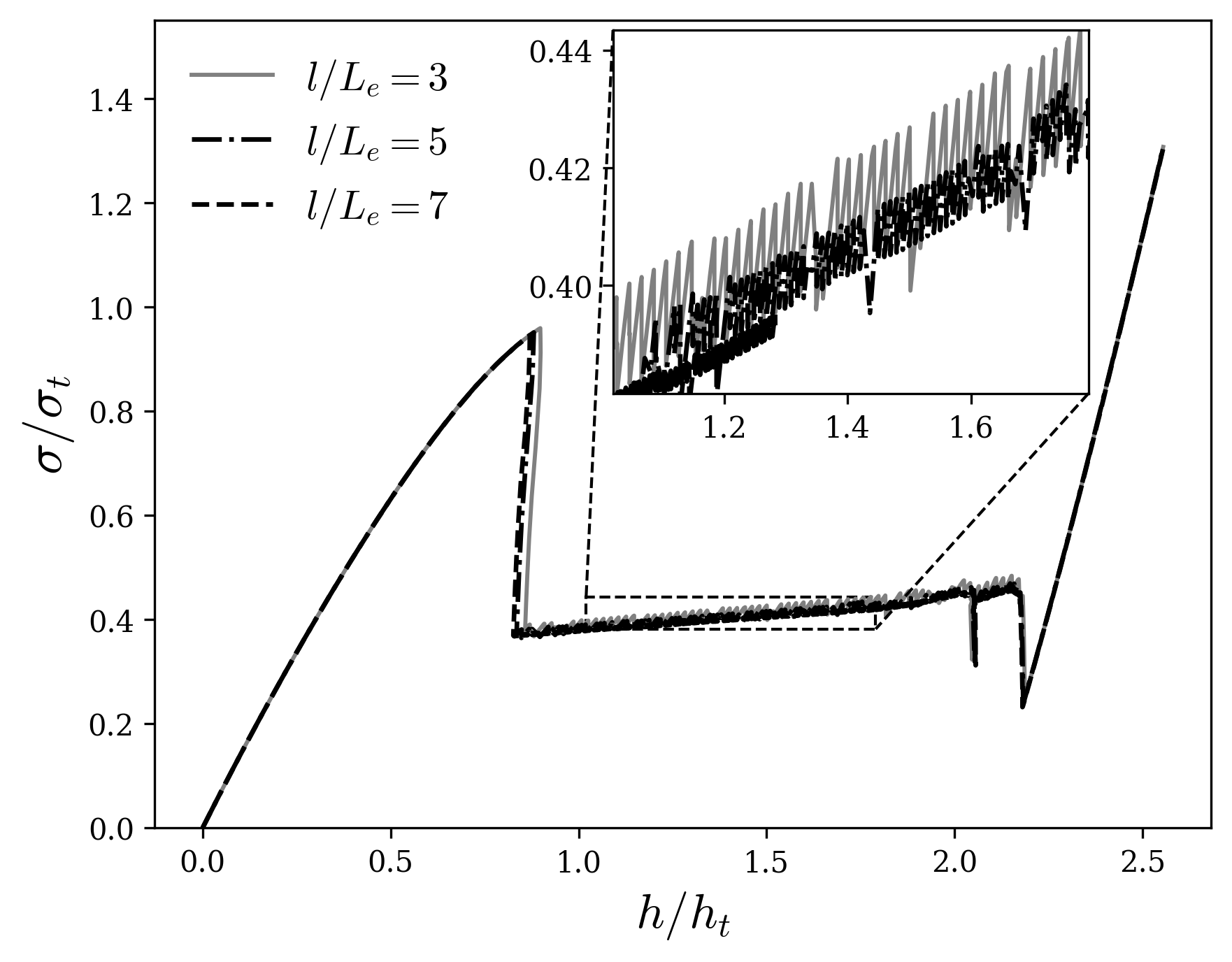}
\caption{Axial, volume-averaged, normalized stress vs strain responses for the strip tension test for mesh densities of $L_e=l/3$, $L_e=l/5$, and $L_e=l/7$ using the local (non–gradient-enhanced), rate-dependent model.}
\label{NoFld}
\end{figure}

It should be noted that the numerical scheme failed to converge throughout the loading process for the rate-independent model, regardless of whether gradient enhancement was included. The values of $\eta$ required to achieve convergence were found to be mesh-size dependent; the smallest the mesh size, the largest the required $\eta$-value. Figure~\ref{eta} compares the uniaxial response of the rate-dependent and rate-independent models using the largest $\eta$-value employed in the simulations, illustrating the extent to which the material response had to be modified to ensure convergence. The strip tension test is also simulated using the local (non–gradient-enhanced), rate-dependent model. As shown in Figure~\ref{NoFld}, the simulated responses exhibit clear mesh dependency. While the peak stress values prior to softening vary with mesh size, the most notable effect is the ``sawtooth pattern" of the stress–strain curve during the propagation of the transition front (\emph{i.e.}, the plateau region). These patterns, whose amplitude also depends on mesh size, arise because the width of the transition front collapses to the size of a single element. As a result, the front propagates through discrete ``jumps" from one element to the next. While rate dependence does not guarantee solution uniqueness or eliminate mesh sensitivity, it is postulated to enhance convergence of the numerical scheme by (i) stabilizing the iterative response through a temporally smooth rather than abrupt evolution of transition localization (\ref{SML}), and (ii) overcoming snap-backs induced by locally non-proportional loading near transition fronts. Additional simulations (not shown) indicate that the required value of $\eta$ for convergence is highly sensitive to the slope and depth of the softening region in the stress–strain curve; for sufficiently shallow and gradual softening, the need for viscous regularization may be entirely eliminated.
\begin{table}[H]
\centering
\caption{Element size, \# of elements, and $\eta$-value needed for numerical convergence for the three meshes used in the tension strip test simulations.}
\label{tab:mesh}
\begin{tabular}{c|c|c|c|c|c}
\toprule
\textbf{Mesh}  & $l/L_e$ & \# of elements & $\eta$ [MPa . s]\\
\midrule
\textbf{Mesh \#1}  & 3 & 6000  & 0.9 \\
\textbf{Mesh \#2}  & 5 & 16650 & 1.3\\
\textbf{Mesh \#3}  & 7 & 32620 & 1.5 \\
\bottomrule
\end{tabular}
\end{table}

\subsection{Radial expansion of a sphere}
\begin {figure}[]
\centering
\subfloat[Geometry and boundary conditions, \emph{i.e.}, prescribed radial displacement during loading and prescribed pressure during unloading.]{\label {schem_sphere}\includegraphics[height=0.25\textheight, keepaspectratio=true]{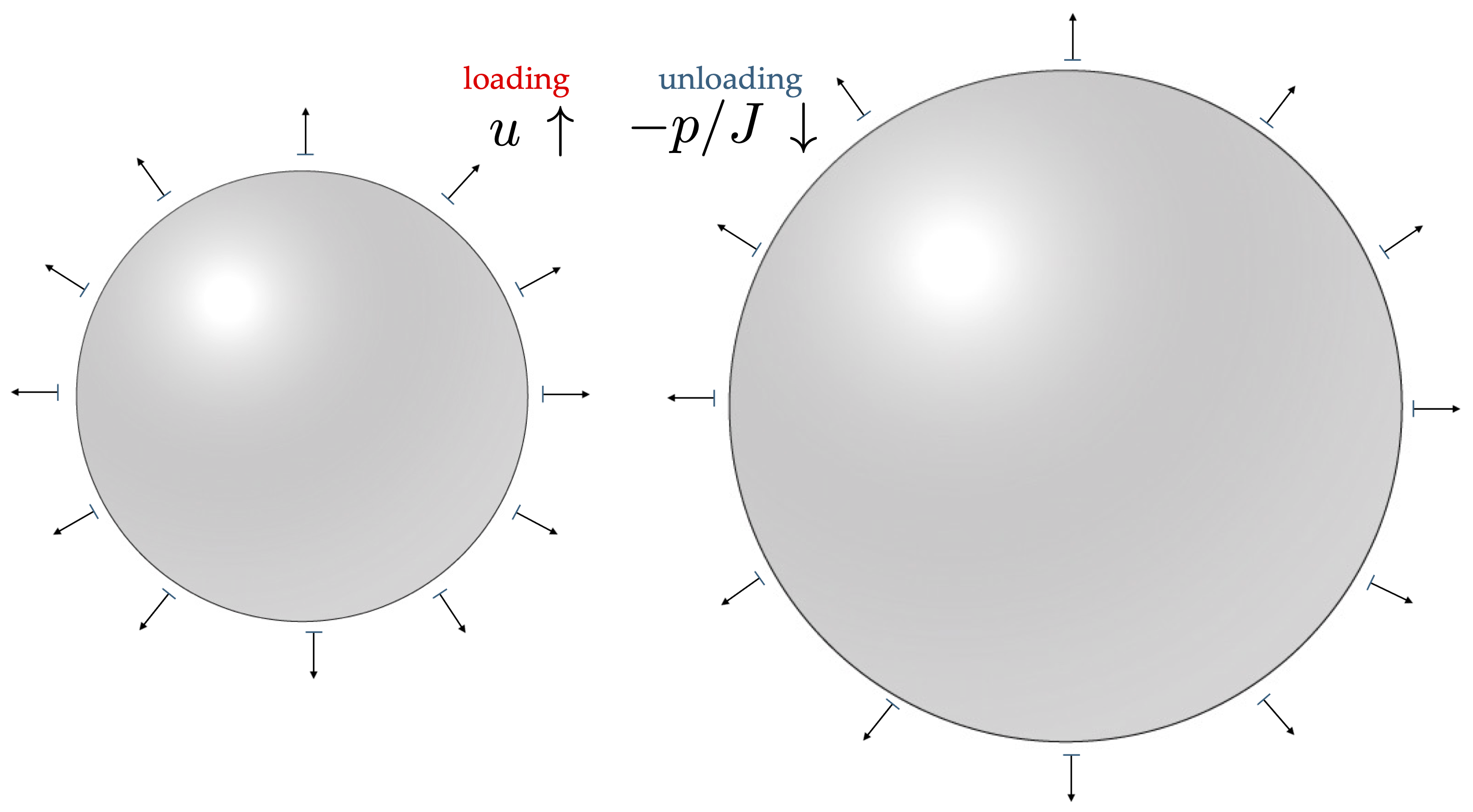}}
\quad
\subfloat[Initial and final stable states.]{\label {IF_sphere}\includegraphics[height=0.25\textheight, keepaspectratio=true]{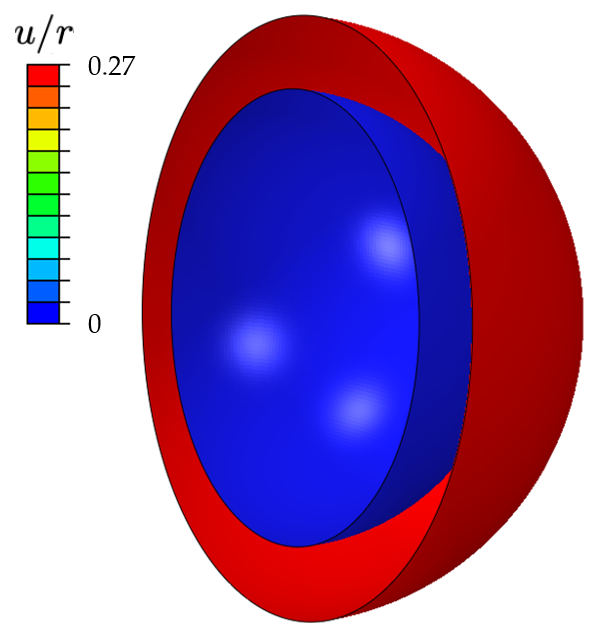}}
\quad
\subfloat[Maximum in-plane stress vs maximum in-plane strain during loading.]{\label {SS_sphere}\includegraphics[width=0.48\textwidth, keepaspectratio=true]{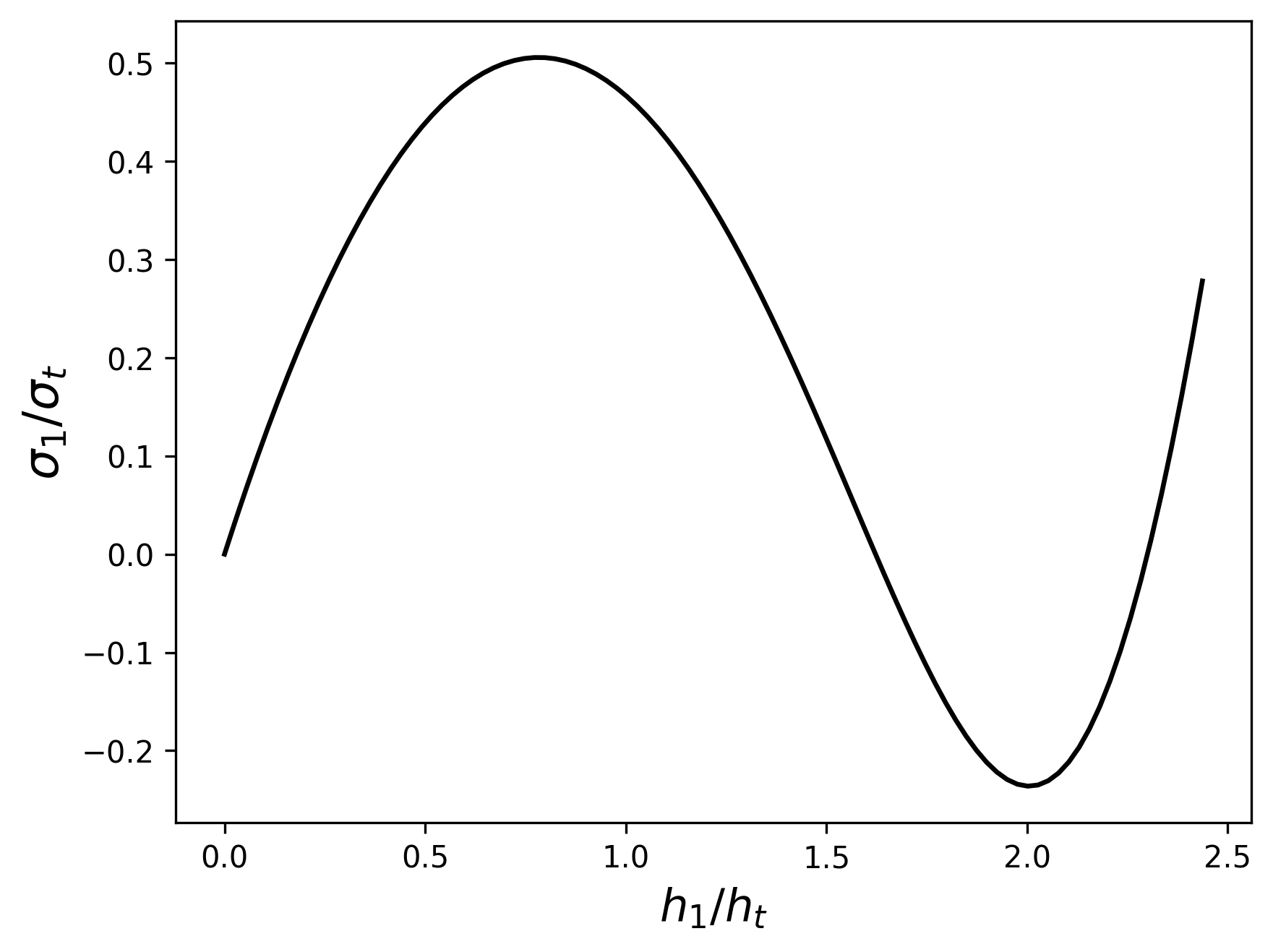}}
\vspace{0pt}
\caption{Radial expansion of a sphere.}
\label {Sphere}
\end {figure}

A hollow sphere with a radius, $r=20$~cm, and thickness, $t=3$~mm,  is subjected to radial outward displacement followed by unloading (Figure~\ref{schem_sphere}). The mesh consists of 40,844 S4 shell elements. The material parameters are given in Table~\ref{tab:mp}. The transverse shear stiffness, calculated based on the shear modulus of the base material as described in Section~\ref{implem}, is $K_{11}=K_{22}=1.0325$~KN/m and $K_{12}=0$~KN/m. Note that no regularization is employed in these simulations. 

The simulation is carried out in two steps. In the first step, the prescribed displacement is gradually increased to its maximum value $u_{max}/r=7/20$ and in a subsequent step the pressure load is progressively reduced to zero. 

In Figure~\ref{IF_sphere}, the initial and final states (the two stable configurations) of half of the sphere are shown. The final configuration (shown in red) exhibits the residual strain characteristic of the second stress minimum of the typical bistable behavior after the pressure load is completely released. he plot of maximum in-plane strain versus maximum in-plane stress on the sphere surface is shown in Figure~\ref{SS_sphere}, and agrees well with the response obtained from equibiaxial loading of a unit cell under periodic boundary conditions.
\begin{sidewaysfigure}
\centering
\includegraphics[width=0.99\textwidth, keepaspectratio=true]{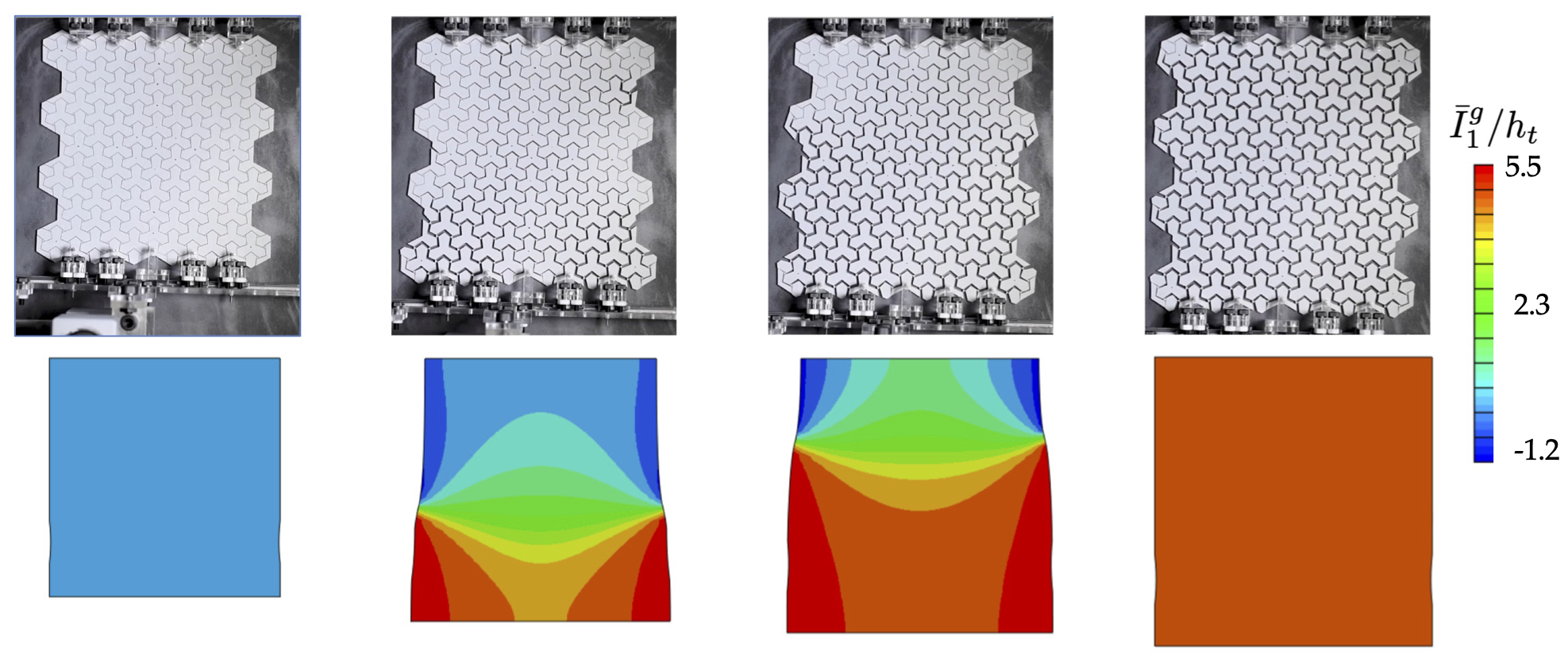}
\caption{Comparison between experimental results and numerical simulations (contours of normalized trace of the strain tensor) for a uniaxial tension experiment. Colors in the experimental results are used for visual guidance in comparing between experiment and simulation.}
\label{Exp}
\end{sidewaysfigure}
\subsection{Uniaxial tension: experiment vs simulation}\label{EV}

In this section, the proposed model is employed to simulate a uniaxial tension experiment on a periodic tessellation of rotating bistable unit cells (Figure~\ref{Exp}). The specimen was fabricated by laser cutting natural rubber sheets (Trodat Aero+). The unit cell geometry (UCGe) parameters and the calibrated material properties of the macro-model, obtained via the procedure described in Section~\ref{CPMM}, are provided in Table~\ref{tab:mpb}, based on the base material properties listed in Table~\ref{tab:mpbm}.

The experiment was conducted using a Saint-Venant-type device~\cite{liu2022triclinic}, which imposes boundary displacement in one direction while allowing free expansion in the other (Figure~\ref{Exp}). Five points on each boundary were connected to grippers via individual carriages mounted on linear bearings. The specimen was supported on a PTFE surface lubricated with talcum powder. Displacement was prescribed using a load testing machine (Instron 68SC-1).

The simulated specimen is a rectangular plate measuring 100 mm $\times$ 95 mm and 2.3 mm thick, uniformly meshed with 1,155 M3D4 membrane elements. A horizontal notch was introduced to trigger state transition at a location that matches the experimental observation. The $\eta$-value used in the simulation is $0.9$ MPa$\cdot$s.

Experimentally, the state transition initiates at unit cells near both ends of the bottom edge and propagates toward the center and upward, eventually spanning the entire specimen (Figure~\ref{Exp}). Discrepancies between experiment and simulation are expected due to boundary condition deviations, fabrication imperfections, and, most notably, the large unit-cell-to-sample-size ratio in the physical specimen, which is in contradiction with the model’s assumption of scale separation. The simulation does capture the qualitative evolution of the transition kinetics, supporting its predictive capability in capturing the deformation response of rotating bistable auxetic surfaces.

\section{Summary}\label{Concl}
A variationally consistent constitutive model was developed to characterize the effective deformation response of periodic rotating auxetic surfaces composed of polymeric hexagonal bistable cells. By expressing the free energy in terms of the invariants of the logarithmic strain tensor, the model enables a physically meaningful decomposition of the Kirchhoff stress into orthogonal components, thereby facilitating straightforward calibration against unit-cell level simulations. To address the mathematical challenges inherent to the double-well energy landscape–such as loss of ellipticity and mesh sensitivity–two regularization strategies were introduced: (i) a nonlocal, gradient-enhanced formulation of the trace of the logarithmic strain tensor and (ii) an artificial viscosity. The gradient-based regularization defines a material length scale that governs the finite width of transition fronts between bistable states, mitigating spurious localization and ensuring mesh-independent, physically realistic finite element responses. The artificial viscosity improves convergence of the nonlinear numerical scheme by promoting a smooth temporal evolution of transition localization and helping the system overcome snap-backs induced by local non-proportional loading near transition fronts. The combined application of gradient-based regularization and artificial viscosity proved effective in capturing the essential features of the deformation response and transition kinetics characteristic of the bistable architecture.

The implementation of the model within the ABAQUS finite element environment using membrane/shell and plane stress elements enables robust simulation of structures comprising these auxetic surfaces. The numerical results confirm the predictive capabilities of the proposed framework, paving the way for its application in the design and analysis of programmable mechanical metamaterials.

Nonetheless, the model has several limitations in its current form:
(i) it is restricted to bistable auxetic surfaces that yield isotropic in-plane response;
(ii) it assumes strain-rate and temperature independence, whereas many polymeric materials exhibit viscoelastic and temperature-dependent behavior; and
(iii) it cannot capture load-controlled scenarios that trigger snap-through instabilities, limiting its applicability to displacement-driven loading conditions.

\section*{Acknowledgements}
This study was supported by the NASA MIRO ``Inflatable Deployable Environments and Adaptive Space Systems'' (IDEAS$^2$) Center under Grant no. 80NSSC24M0178. The authors acknowledge the \href{https://www.uh.edu/rcdc/}{Research Computing Data Core (RCDC)} at the University of Houston for the supercomputing resources made available for conducting the research reported in this paper.

\appendix
\section{A Simplified Model for Domain Transition Localization}\label{SML}

The fundamental mechanism driving a domain localization can be captured by a simple bar model (Figure~\ref{strip_sch}), representing a one-dimensional approximation of a long plate strip under tension~\cite{hutchinson1977influence, 10.1115/1.3153742,needleman1988material}. Linearized kinematics are assumed; although the extension to finite strains is technically straight-forward, no new physics are expected, and linearized kinematics are preferred for illustrative simplicity.

\subsection{Rate-independent model}\label{RIC}
\begin{figure}
\centering
\includegraphics[width=0.4\textwidth, keepaspectratio=true]{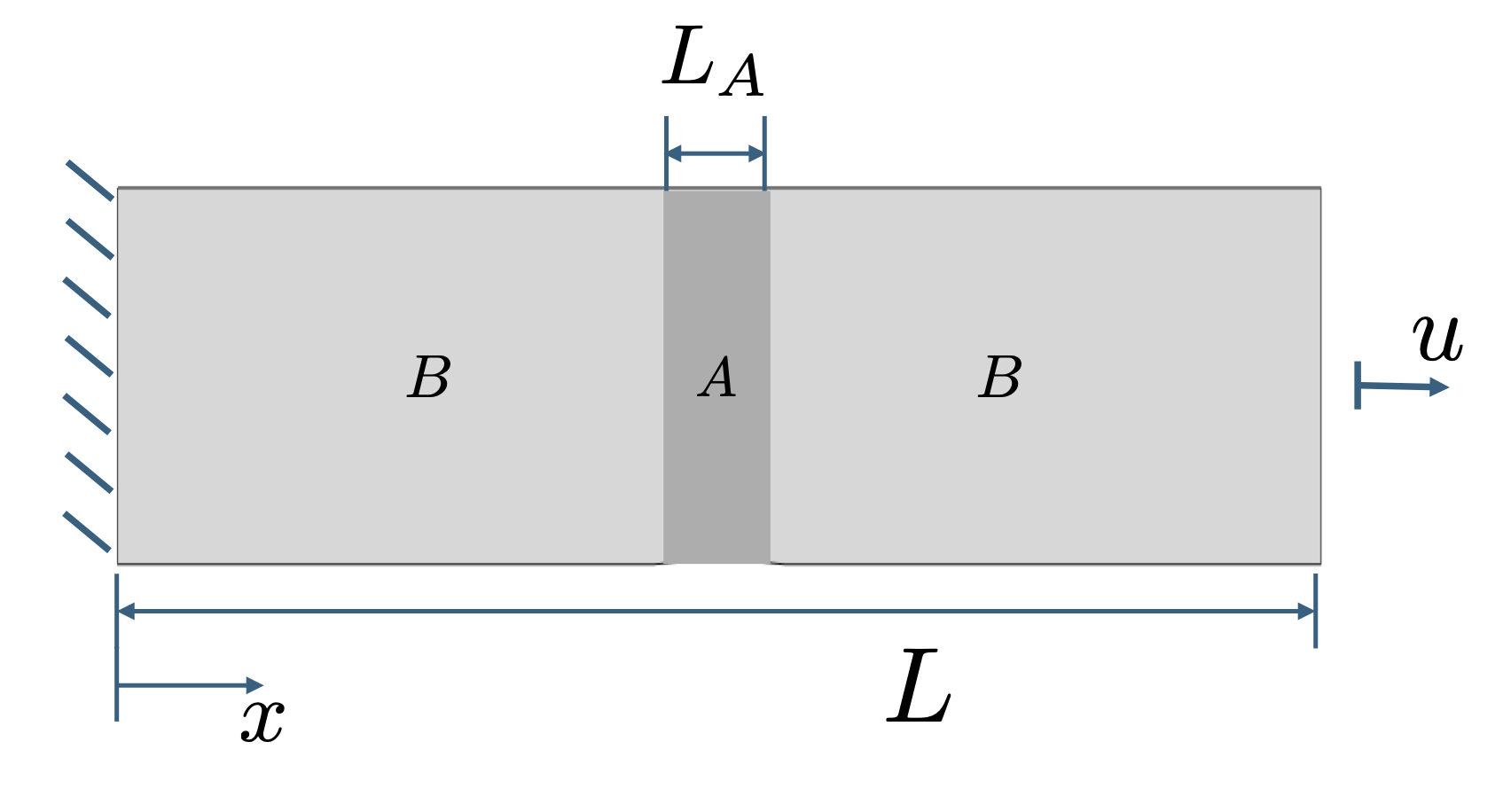}
\caption{Uniaxial tension of a planar strip. The only spatial dependence of field quantities is on $x$. Also shown is a
A planar band of thickness $L_A$ and relative thickness $\alpha=L_A/L$.}
\label{strip_sch}
\end{figure}

The incremental 1D relation between axial stress, $\sigma$, and strain, $\varepsilon$, derived from~\eqref{hypercl}, assuming small strains, is
\begin{equation}\label{cr}
\delta{\sigma}=\left(\frac{\partial \psi}{\partial \varepsilon}\right)_{\varepsilon}\delta{\varepsilon},
\end{equation}
where $\delta$ denotes an increment and $()_{\varepsilon}$ differentiation with respect to $\varepsilon$. At some stage of the deformation history, the possibility of bifurcation into a localized region (region A in Figure~\ref{strip_sch}), which undergoes incremental straining different from that in the surrounding material, is considered. Neglecting possible normal strains or normal stresses due to nonlinear kinematic effects, equilibrium requires that the stress state remain homogeneous so that
\begin{equation}\label{ie}
\delta{\sigma}_A=\delta{\sigma}_B,
\end{equation}
where the subscripts $A$ and $B$ denote quantities associated with the corresponding regions in the figure. Thus, on account of~\eqref{cr}
\begin{equation}
\left(\frac{\partial \psi}{\partial \varepsilon}\right)_{\varepsilon}\left(\delta{\varepsilon}_A-\delta{\varepsilon}_B\right)=0,
\end{equation}
which implies that such an alternative deformation state is possible only when $\left(\frac{\partial \psi}{\partial \varepsilon}\right)_{\varepsilon}=0$, \emph{i.e.}, at the maximum load point. However, it is important to emphasize that in a more general multi-axial setting, strain softening is neither a necessary nor a sufficient condition for localization. 

\begin{figure}
\centering
\includegraphics[width=0.35\textwidth, keepaspectratio=true]{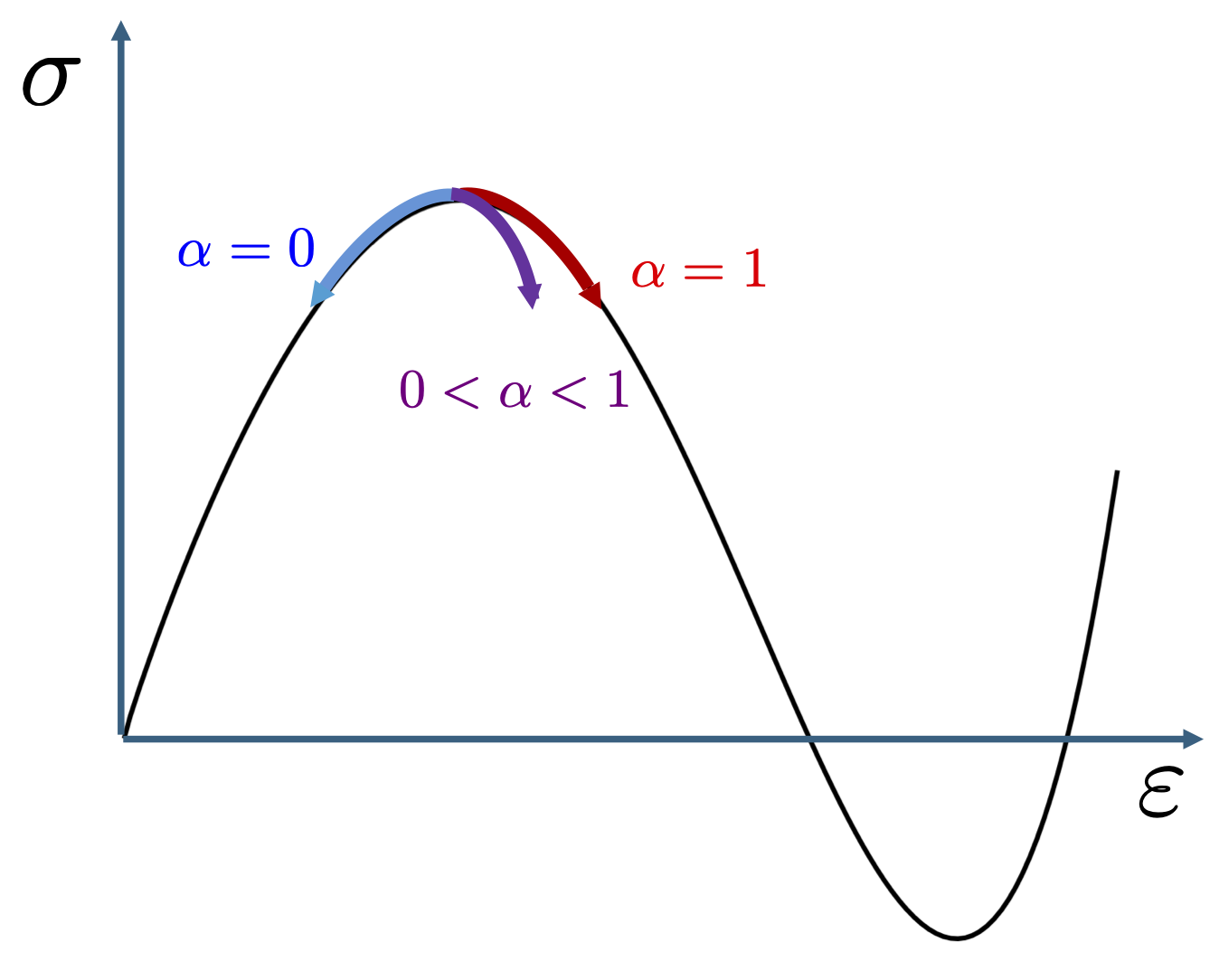}
\caption{The dependence of the overall stress vs strain response on the relative thickness of the localization band $\alpha=L_A/L$ for the rate-independent case.}
\label{alpha}
\end{figure}

In the post-bifurcation regime, the total nominal strain increment is expressed as
\begin{equation}
\delta{\varepsilon}=(1-\alpha)\delta{\varepsilon}_A+\alpha \delta{\varepsilon}_B,
\end{equation}
where $\alpha=L_A/L$, and, on account of \eqref{ie} and~\eqref{cr},
\begin{equation}
\delta{\varepsilon}=\frac{\delta{\sigma}}{C_A}\left[\alpha+\left(1-\alpha\right)\frac{C_A}{C_B}\right],
\end{equation}
where for convenience in notation $\left(\frac{\partial \psi}{\partial \varepsilon}\right)_{\varepsilon}$, is replaced by $C$. At bifurcation, \emph{i.e.}, at the maximum load, $C_B = C_A=C^b=0$. Just after bifurcation occurs, the stress drops (in both regions); the material inside the band is further stretched while the material outside the band is elastically unloading (if the material outside the band were to continue stretching, the strain increments in the two regions would be identical and there would be no band). The stress–strain response associated with localization lies below that corresponding to homogeneous deformation; for a thick band ($\alpha$  close to unity), the overall response follows the loading curve of the homogeneous material, while for a sufficiently thin band ($\alpha$ near zero), the overall response approximates the elastic unloading branch (Figure~\ref{alpha}). 

The value of $\alpha$ remains undetermined by the analysis, reflecting the inherent non-uniqueness of the solution. Notably, as discussed in~\cite{needleman1988material}, the presence of an initial imperfection does not resolve this non-uniqueness. When an imperfection, in the form of a front, is introduced, the maximum stress within the band is reached prior to that in the surrounding material. Once the maximum stress is attained in the band, the bifurcation arguments outlined above apply, with the bar length effectively replaced by the length of the imperfection band. Thus, localization, \emph{i.e.}, loss of ellipticity, inevitably gives rise to inherently non-unique solutions characterized by bifurcation modes in the form of arbitrarily narrow fronts.
%

\subsection{Rate-dependent model}\label{RIC}
A similar stability analysis can be extended to the rate-dependent regularized formulation. A perturbation about the homogeneous base state, denoted by $(\delta\sigma,\, \delta\varepsilon)$, satisfies
\begin{equation}\label{rdms}
\delta\sigma = \left(\frac{\partial \psi}{\partial \varepsilon}\right)_{\varepsilon}\delta\varepsilon + \eta\,\delta\dot{\varepsilon}.
\end{equation}
Under quasi-static equilibrium, $\delta\sigma = 0$. Assuming a Fourier-mode perturbation $\delta\varepsilon(x,t) = \hat{\varepsilon}\,e^{ikx + st}$, where $k \neq 0$ is the wavenumber and $s$ is the temporal growth rate, differentiation with respect to time yields
\begin{equation}
\left[\left(\frac{\partial \psi}{\partial \varepsilon}\right)_{\varepsilon} + \eta s\right] \delta \varepsilon = 0,
\end{equation}
which, for non-trivial perturbations $(\delta\varepsilon \neq 0)$, leads to
\begin{equation}
s = -\frac{\left(\frac{\partial \psi}{\partial \varepsilon}\right)_{\varepsilon}}{\eta}.
\end{equation}

The temporal character of the perturbation depends on the sign of $\left(\frac{\partial \psi}{\partial \varepsilon}\right)_{\varepsilon}$:
\begin{itemize}
\item $\left(\frac{\partial \psi}{\partial \varepsilon}\right)_{\varepsilon} > 0$: $s < 0$, perturbations decay exponentially in time (stable regime);
\item $\left(\frac{\partial \psi}{\partial \varepsilon}\right)_{\varepsilon} = 0$: $s = 0$, neutral equilibrium;
\item $\left(\frac{\partial \psi}{\partial \varepsilon}\right)_{\varepsilon} < 0$: $s > 0$, perturbations grow exponentially (unstable regime).
\end{itemize}

Hence, material softening characterized by $\left(\frac{\partial \psi}{\partial \varepsilon}\right)_{\varepsilon} < 0$ leads to temporal instability in the form of localization.

\subsection{Gradient-enhanced rate-dependent model}\label{GRIC}

For the one-dimensional gradient-enhanced rate-dependent model,
\begin{equation}
\begin{cases}
\sigma = \dfrac{\partial \psi}{\partial \bar{\varepsilon}} + \eta\,\dot{\varepsilon}, \\
\bar{\varepsilon} - l^{2}\bar{\varepsilon}_{xx} = \varepsilon,
\end{cases}
\end{equation}
small perturbations $(\delta\sigma,\, \delta\varepsilon,\, \delta\bar{\varepsilon})$ about a homogeneous reference state satisfy
\begin{equation}\label{abaa}
\begin{cases}
\delta\sigma = \left(\dfrac{\partial \psi}{\partial \bar{\varepsilon}}\right)_{\bar{\varepsilon}}\,\delta\bar{\varepsilon} + \eta\,\delta\dot{\varepsilon}, \\
\delta\bar{\varepsilon} - l^{2}\delta\bar{\varepsilon}_{xx} = \delta\varepsilon.
\end{cases}
\end{equation}

Assuming perturbations of the form
\begin{equation}
(\delta\varepsilon,\, \delta\bar{\varepsilon}) \propto e^{ikx+st}, \quad k \neq 0,
\end{equation}
the Helmholtz-type relation \eqref{abaa}$_2$ yields
\begin{equation}
\delta\bar{\varepsilon} = \frac{\delta\varepsilon}{1 + l^{2}k^{2}}.
\end{equation}
Enforcing equilibrium $\delta\sigma = 0$,  \eqref{abaa}$_1$ gives
\begin{equation}
\left[\frac{\left(\dfrac{\partial \psi}{\partial \bar{\varepsilon}}\right)_{\bar{\varepsilon}}}{1 + l^{2}k^{2}} + \eta s\right]\delta\varepsilon = 0.
\end{equation}
For nontrivial perturbations $(\delta\varepsilon \neq 0)$, the dispersion relation—relating the temporal growth rate to wavelength—becomes
\begin{equation}
s(k) = -\,\frac{\left(\dfrac{\partial \psi}{\partial \bar{\varepsilon}}\right)_{\bar{\varepsilon}}}{\eta\,(1 + l^{2}k^{2})}.
\end{equation}

The stability characteristics follow as:
\begin{itemize}
\item $\left(\dfrac{\partial \psi}{\partial \bar{\varepsilon}}\right)_{\bar{\varepsilon}} > 0$: $s(k) < 0$, perturbations decay exponentially (stable);
\item $\left(\dfrac{\partial \psi}{\partial \bar{\varepsilon}}\right)_{\bar{\varepsilon}} = 0$: $s(k) = 0$, neutral equilibrium;
\item $\left(\dfrac{\partial \psi}{\partial \bar{\varepsilon}}\right)_{\bar{\varepsilon}} < 0$: $s(k) > 0$, perturbations grow exponentially (unstable).
\end{itemize}

The magnitude of the rate $|s(k)|$ decreases monotonically with wavenumber $k$. Thus, short-wavelength disturbances (large $k$) are suppressed since $s(k) \to 0$ as $k \to \infty$. For $\left(\frac{\partial \psi}{\partial \bar{\varepsilon}}\right)_{\bar{\varepsilon}} < 0$, the maximum growth rate occurs at the long-wave limit $k_{\max} = 0$, with $s_{\max} = |\left(\frac{\partial \psi}{\partial \bar{\varepsilon}}\right)_{\bar{\varepsilon}} |/\eta$. Therefore, in the linear regime the model does not select a finite preferred wavelength but promotes diffuse, long-wave amplification. Nonetheless, the Helmholtz relation ensures that any spatial variation in $\delta \varepsilon$ produces a smoothed $\delta \bar{\varepsilon}$ over a characteristic length scale $\sim l$, so localization in the nonlinear regime attains a finite width on the order of $l$.

\section{Constitutive Response of the Base Material}\label{CRBM}
The base material in the unit cell simulations used for the calibration of the continuum macro-model is natural rubber, modeled via a neo-Hookean hyperelastic model.

The strain energy density function of a neo-Hookean hyperelastic material is given as 
\begin{equation}
U\left(\hat{I}_{1}, \hat{I}_{2}, J\right)=C_{10}\left(\hat{I}_{1}-3\right)+\frac{1}{D_{1}}(J-1)^{2} ,
\end{equation}
where $\hat{I}_{1}= \operatorname{tr}\left(\bar{\boldsymbol{B}}\right)$, $\hat{I}_{2}=\frac{1}{2}\left[\hat{I}_{1}^2-\operatorname{tr}\left(\bar{\boldsymbol{B}}^2\right)\right]$ are strain invariants of the deviatoric left Cauchy-Green deformation tensor $\bar{\boldsymbol{B}}=J^{-\frac{1}{3}}\boldsymbol{F} J^{-\frac{1}{3}}\boldsymbol{F}^T$ in which the volume change has been eliminated.
Hence, the constitutive equation for a neo-Hookean material reads as
\begin{equation}
\boldsymbol{\sigma}=\frac{2}{J} C_{10}\left[\bar{\boldsymbol{B}}-\frac{1}{3} \operatorname{tr}\left(\bar{\boldsymbol{B}}\right) \boldsymbol{\delta}\right]+\frac{2}{D_{1}}(J-1) \boldsymbol{\delta}.
\end{equation}
 $C_{10}=\mu_{0}/2$ and $D_{1}=2/K_{0}$, where $\mu_{0}$ is the initial shear modulus and $K_{0}$ the initial bulk modulus, are material parameters that describe the shear and volumetric material response, respectively. For typical natural rubber, these parameters take the values listed in Table~\ref{tab:mpbm}, which correspond to an initial Poisson ratio of 0.49991.
\begin{table}[H]
\centering
\caption{Material parameters for neo-Hookean hyperelastic model for natural rubber~\cite{kelly2011use}.}
\label{tab:mpbm}
\renewcommand{\arraystretch}{1.2}
			\begin{tabular}{l|r} \toprule
			parameter & value \\ \midrule
			$\mu_{0}$ [MN/m$^{2}$] & 0.413 \\
			$K_{0}$ [MN/m$^{2}$] & 2300 \\
			\bottomrule
                          \end{tabular}	
\end{table}

\section{Calibration of the Proposed Macro-Model}\label{CPMM}
A heuristic procedure for the calibration of the proposed model is outlined below.

The calibration is performed using data from uniaxial loading of the unit cell under periodic boundary conditions, as discussed in Section~\ref{cal}. In terms of the principal stretches, $\lambda_1 \ge \lambda_2$, and the first Piola-Kirchhoff stress tensor, $\boldsymbol{P} = \boldsymbol{\tau} \boldsymbol{F}^{-T}$, the constitutive equations~\eqref{hypercl}, under uniaxial loading, read as
\begin{align}\label{al}
\begin{dcases}
\lambda_1 P_{1} &= \frac{\partial \psi}{\partial \bar{I}_1} + \operatorname{sgn}(\lambda_1-\lambda_2) \frac{\sqrt{2}}{2} \frac{\partial \psi}{\partial \bar{I}_2} \\
0 &= \frac{\partial \psi}{\partial \bar{I}_1} - \operatorname{sgn}(\lambda_1-\lambda_2) \frac{\sqrt{2}}{2} \frac{\partial \psi}{\partial \bar{I}_2}
\end{dcases}
\Longrightarrow
\begin{dcases}
\frac{\lambda_1 P_{1}}{2} = \frac{\partial \psi}{\partial \bar{I}_1}\\
\frac{\lambda_1 P_{1}}{2} =  \operatorname{sgn}(\lambda_1-\lambda_2) \frac{\sqrt{2}}{2} \frac{\partial \psi}{\partial \bar{I}_2},
\end{dcases}
\end{align}
where $\bar{I}_1 = \ln \lambda_1 + \ln \lambda_2$, $\bar{I}_2 = \sqrt{2} \left| \ln \lambda_1 - \ln \lambda_2 \right|/2$, and 
\[
\operatorname{sgn}(\lambda_1-\lambda_2)=
\begin{cases}
-1, & \lambda_1< \lambda_2,\\
0, & \lambda_1= \lambda_2, \\
1, & \lambda_1> \lambda_2,
\end{cases}
\]
is the sign (or signum) function.

\begin {figure}[]
\centering
\subfloat[Calibration of the parameters $a_i$ ($i=1...5$) related to the volumetric deformation response.]{\label {BC}\includegraphics[width=0.48\textwidth, keepaspectratio=true]{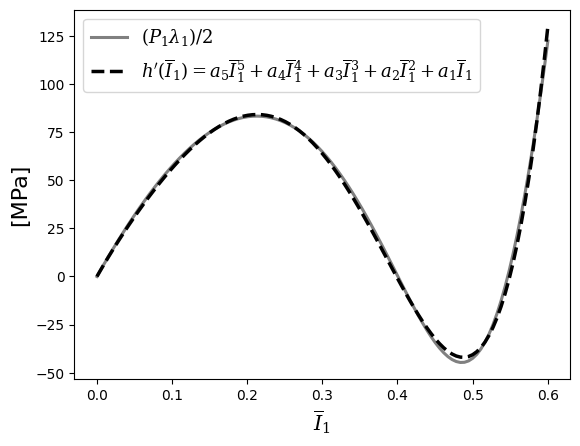}}
\quad
\subfloat[Calibration of the parameters $c$ related to the deviatoric (\emph{i.e.}, volume preserving) deformation response.]{\label {UC}\includegraphics[width=0.48\textwidth, keepaspectratio=true]{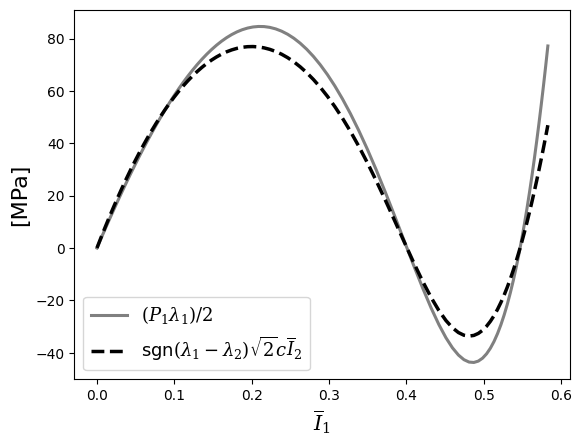}}
\vspace{0pt}
\caption{Model calibration from uniaxial stretching of a unit cell under periodic boundary conditions.}
\label {BCuc}
\end {figure}

\begin{figure}
\centering
\includegraphics[height=0.4\textheight, keepaspectratio=true]{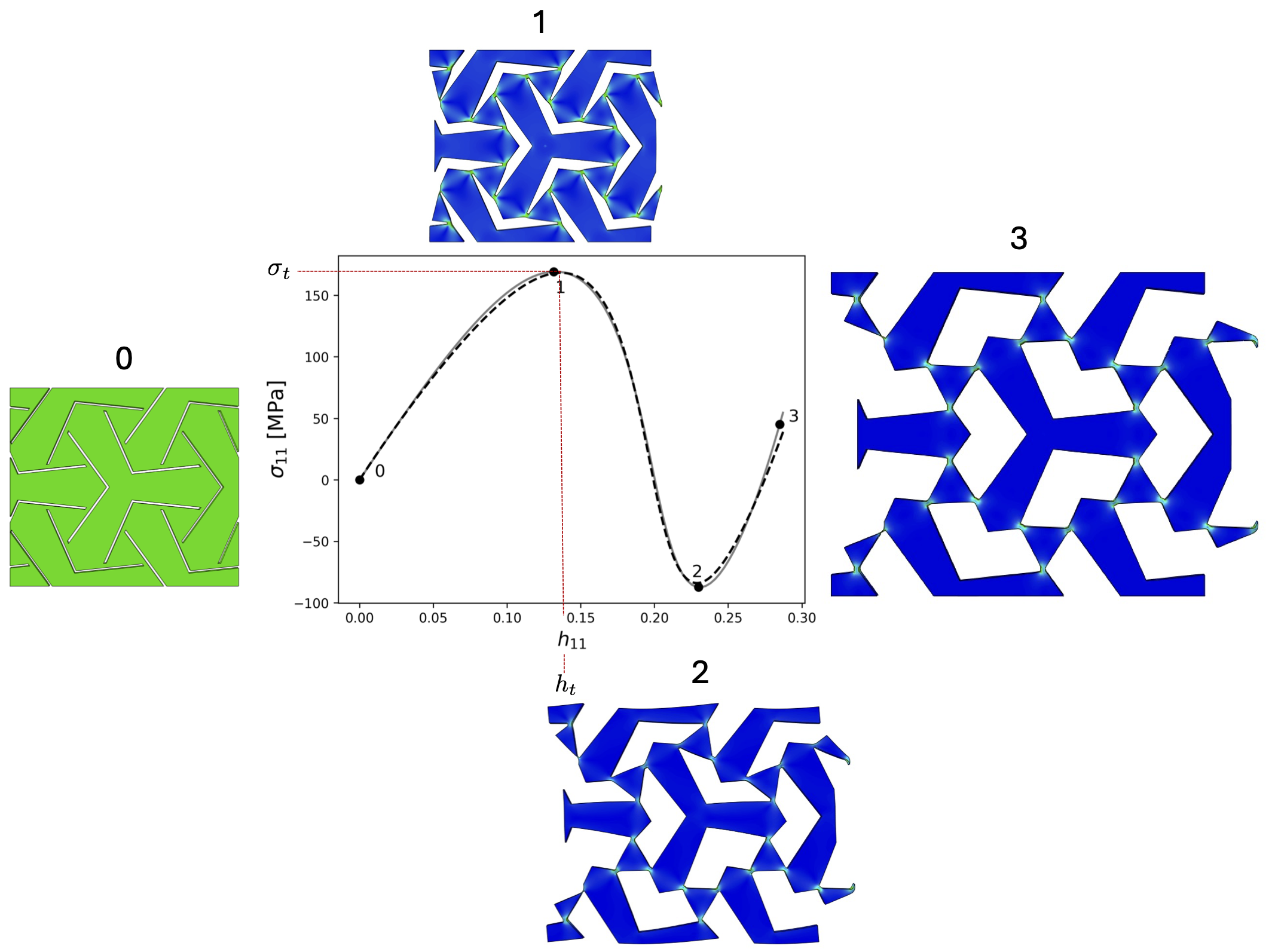}
\caption{Uniaxial stress vs strain response of the unit cell UCG characterized by the pattern parameters listed in Table~\ref{tab:mpa}. $\sigma_t$ and $h_t$ are the stress and strain values used in the stress and strain normalization adopted in Section~\ref{tension}, respectively.}
\label{Dots}
\end{figure}

The left-hand sides of the above equations are obtained from the unit cell simulations, while the right-hand sides are least-square fitted, assuming
\begin{equation}
\frac{\partial \psi}{\partial \bar{I}_1} = h’(\bar{I}_1) = a_5 \bar{I}_1^5 + a_4 \bar{I}_1^4 + a_3 \bar{I}_1^3 + a_2 \bar{I}_1^2 + a_1 \bar{I}_1,
\end{equation}
and
\begin{equation}
\frac{\partial \psi}{\partial \bar{I}_2} = 2c \bar{I}_2,
\end{equation}
where $a_i$ ($i=1,\dots,5$) and $c$ are model constants. Fifth order polynomials with just two extremum points are chosen for a better fit of the data. The calibration is depicted in Figures~\ref{BC} and~\ref{UC} for the unit cell geometry UCG characterized by the pattern parameters listed in Table~\ref{tab:mpa} (Figure~\ref{triangle}). The material parameter values obtained are also listed in the same table. Furthermore, the uniaxial response obtained from the calibrated model is compared against the uniaxial response of the unit cell under periodic boundary conditions in Figure~\ref{Dots}.

\begin{figure}
\centering
\includegraphics[height=0.4\textheight, keepaspectratio=true]{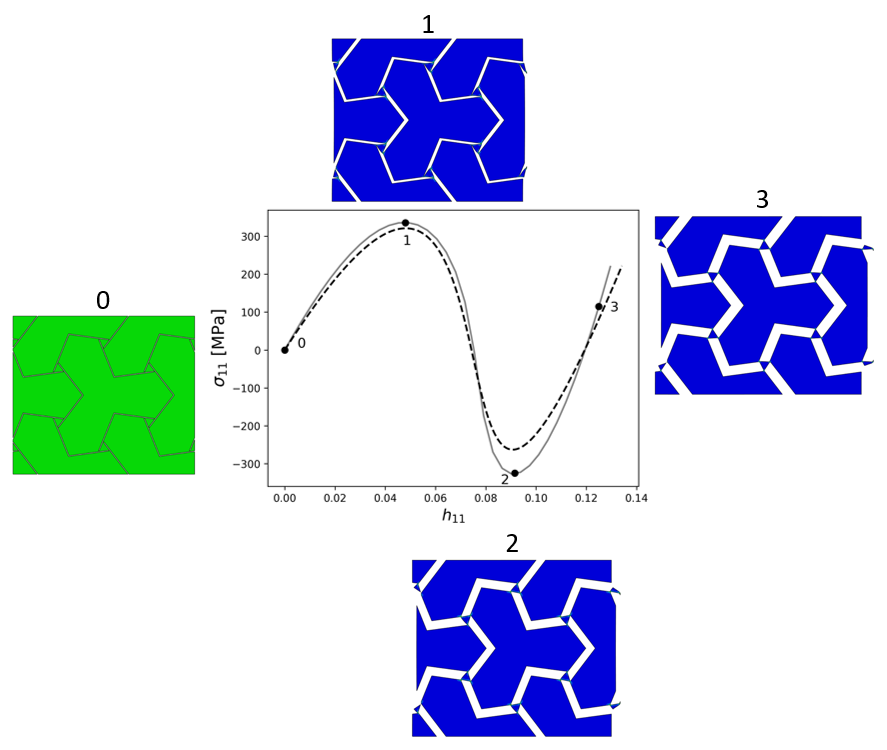}
\caption{Uniaxial stress vs strain response of the unit cell UCGe used in the experiment discussed in Section~\ref{EV}.}
\label{Exper_UC}
\end{figure}

In Table~\ref{tab:mpb}, the calibrated material parameter values for another unit cell geometry UCGe are listed along with the pattern parameters characterizing the geometry. This is the unit cell geometry used in the experiment simulated in Section~\ref{EV} and is shown in Figure~\ref{Exper_UC}.

\begin{table}[H]
	\centering
	\small{	
	\caption{Material parameters for the proposed 2D macro-model calibrated for the two unit cell geometries UCG and UCGe for the base material parameters listed in Table~\ref{tab:mpbm}.}
\label{tab:mp}

\subfloat[Unit Cell Geometry UCG \label{tab:mpa}]{
			\renewcommand{\arraystretch}{1}
			\vspace{-10pt}
			\centering
			\begin{tabular}{ llllll} \toprule
			\multicolumn{6}{l}{{\bf Pattern Parameters}}	 \\ \midrule
		  $l_{0}=5.5$~mm & $b=0.77$~mm & $s=0.125$~mm & $w=0.25$~mm & $\theta=6^{\circ}$ &\\ \toprule
			\multicolumn{6}{l}{{\bf Calibrated Material Parameters [MPa]}}	 \\ \midrule
		  $a_5 = 38960$ & $a_4= -30256$ & $a_3 = 5325.7$ & $a_2 = -1502.8$ & $a_1 = 685.66$ & $c= 1408.4$ \\
			\bottomrule
                          \end{tabular}	}
                         
                          \vspace{5pt}
                          
                          \subfloat[Unit Cell Geometry UCGe  \label{tab:mpb}]{
			\renewcommand{\arraystretch}{1}
			\vspace{-10pt}
			\centering
			\begin{tabular}{ llllll} \toprule
			\multicolumn{6}{l}{{\bf Pattern Parameters}}	 \\ \midrule
		  $l_{0}=15$~mm & $b=3.359$~mm & $s=0.125$~mm & $w=0.15$~mm & $\theta=7.72^{\circ}$ & (thickness $\sim$ 2.3~ mm)\\ \toprule
			\multicolumn{6}{l}{{\bf Calibrated Material Parameters [MPa]}}	 \\ \midrule
		  $a_5 = 3000000$ & $a_4= -302937$ & $a_3 = -116403$ & $a_2 = -12106$ & $a_1 = 3901.1$ & $c= 7662.77$ \\
			\bottomrule
                          \end{tabular}	}}

\end{table}

\bibliography{article}

\begin{thebibliography}{73}
\expandafter\ifx\csname natexlab\endcsname\relax\def\natexlab#1{#1}\fi
\providecommand{\url}[1]{\texttt{#1}}
\providecommand{\href}[2]{#2}
\providecommand{\path}[1]{#1}
\providecommand{\DOIprefix}{doi:}
\providecommand{\ArXivprefix}{arXiv:}
\providecommand{\URLprefix}{URL: }
\providecommand{\Pubmedprefix}{pmid:}
\providecommand{\doi}[1]{\href{http://dx.doi.org/#1}{\path{#1}}}
\providecommand{\Pubmed}[1]{\href{pmid:#1}{\path{#1}}}
\providecommand{\bibinfo}[2]{#2}
\ifx\xfnm\relax \def\xfnm[#1]{\unskip,\space#1}\fi
\bibitem[{Wang and Hu(2014)}]{wang2014auxetic}
\bibinfo{author}{Z.~Wang}, \bibinfo{author}{H.~Hu},
\newblock \bibinfo{title}{Auxetic materials and their potential applications in
  textiles},
\newblock \bibinfo{journal}{Textile Research Journal} \bibinfo{volume}{84}
  (\bibinfo{year}{2014}) \bibinfo{pages}{1600--1611}.
\bibitem[{Sanami et~al.(2014)Sanami, Ravirala, Alderson, and
  Alderson}]{sanami2014auxetic}
\bibinfo{author}{M.~Sanami}, \bibinfo{author}{N.~Ravirala},
  \bibinfo{author}{K.~Alderson}, \bibinfo{author}{A.~Alderson},
\newblock \bibinfo{title}{Auxetic materials for sports applications},
\newblock \bibinfo{journal}{Procedia Engineering} \bibinfo{volume}{72}
  (\bibinfo{year}{2014}) \bibinfo{pages}{453--458}.
\bibitem[{Mir et~al.(2014)Mir, Ali, Sami, and Ansari}]{mir2014review}
\bibinfo{author}{M.~Mir}, \bibinfo{author}{M.~N. Ali},
  \bibinfo{author}{J.~Sami}, \bibinfo{author}{U.~Ansari},
\newblock \bibinfo{title}{Review of mechanics and applications of auxetic
  structures},
\newblock \bibinfo{journal}{Advances in Materials Science and Engineering}
  \bibinfo{volume}{2014} (\bibinfo{year}{2014}) \bibinfo{pages}{753496}.
\bibitem[{Carneiro et~al.(2013)Carneiro, Meireles, and
  Puga}]{carneiro2013auxetic}
\bibinfo{author}{V.~H. Carneiro}, \bibinfo{author}{J.~Meireles},
  \bibinfo{author}{H.~Puga},
\newblock \bibinfo{title}{Auxetic materials—a review},
\newblock \bibinfo{journal}{Materials Science-Poland} \bibinfo{volume}{31}
  (\bibinfo{year}{2013}) \bibinfo{pages}{561--571}.
\bibitem[{Yang et~al.(2004)Yang, Li, Shi, Xie, and Yang}]{yang2004review}
\bibinfo{author}{W.~Yang}, \bibinfo{author}{Z.-M. Li},
  \bibinfo{author}{W.~Shi}, \bibinfo{author}{B.-H. Xie}, \bibinfo{author}{M.-B.
  Yang},
\newblock \bibinfo{title}{Review on auxetic materials},
\newblock \bibinfo{journal}{Journal of materials science} \bibinfo{volume}{39}
  (\bibinfo{year}{2004}) \bibinfo{pages}{3269--3279}.
\bibitem[{Ren et~al.(2018)Ren, Das, Tran, Ngo, and Xie}]{ren2018auxetic}
\bibinfo{author}{X.~Ren}, \bibinfo{author}{R.~Das}, \bibinfo{author}{P.~Tran},
  \bibinfo{author}{T.~D. Ngo}, \bibinfo{author}{Y.~M. Xie},
\newblock \bibinfo{title}{Auxetic metamaterials and structures: a review},
\newblock \bibinfo{journal}{Smart materials and structures}
  \bibinfo{volume}{27} (\bibinfo{year}{2018}) \bibinfo{pages}{023001}.
\bibitem[{Lim(2015)}]{lim2015auxetic}
\bibinfo{author}{T.-C. Lim}, \bibinfo{title}{Auxetic materials and structures},
  volume~\bibinfo{volume}{34}, \bibinfo{publisher}{Springer},
  \bibinfo{year}{2015}.
\bibitem[{Scarpa(2008)}]{scarpa2008auxetic}
\bibinfo{author}{F.~Scarpa},
\newblock \bibinfo{title}{Auxetic materials for bioprostheses [in the
  spotlight]},
\newblock \bibinfo{journal}{IEEE Signal Processing Magazine}
  \bibinfo{volume}{25} (\bibinfo{year}{2008}) \bibinfo{pages}{128--126}.
\bibitem[{Jiang and Li(2018)}]{jiang20183d}
\bibinfo{author}{Y.~Jiang}, \bibinfo{author}{Y.~Li},
\newblock \bibinfo{title}{3d printed auxetic mechanical metamaterial with
  chiral cells and re-entrant cores},
\newblock \bibinfo{journal}{Scientific reports} \bibinfo{volume}{8}
  (\bibinfo{year}{2018}) \bibinfo{pages}{1--11}.
\bibitem[{Bettini et~al.(2010)Bettini, Airoldi, Sala, Di~Landro, Ruzzene, and
  Spadoni}]{bettini2010composite}
\bibinfo{author}{P.~Bettini}, \bibinfo{author}{A.~Airoldi},
  \bibinfo{author}{G.~Sala}, \bibinfo{author}{L.~Di~Landro},
  \bibinfo{author}{M.~Ruzzene}, \bibinfo{author}{A.~Spadoni},
\newblock \bibinfo{title}{Composite chiral structures for morphing airfoils:
  Numerical analyses and development of a manufacturing process},
\newblock \bibinfo{journal}{Composites Part B: Engineering}
  \bibinfo{volume}{41} (\bibinfo{year}{2010}) \bibinfo{pages}{133--147}.
\bibitem[{Jayanty et~al.(2011)Jayanty, Crowe, and Berhan}]{jayanty2011auxetic}
\bibinfo{author}{S.~Jayanty}, \bibinfo{author}{J.~Crowe},
  \bibinfo{author}{L.~Berhan},
\newblock \bibinfo{title}{Auxetic fibre networks and their composites},
\newblock \bibinfo{journal}{physica status solidi (b)} \bibinfo{volume}{248}
  (\bibinfo{year}{2011}) \bibinfo{pages}{73--81}.
\bibitem[{Chi et~al.(2022)Chi, Li, Zhao, Hong, Tang, and Yin}]{chi2022bistable}
\bibinfo{author}{Y.~Chi}, \bibinfo{author}{Y.~Li}, \bibinfo{author}{Y.~Zhao},
  \bibinfo{author}{Y.~Hong}, \bibinfo{author}{Y.~Tang},
  \bibinfo{author}{J.~Yin},
\newblock \bibinfo{title}{Bistable and multistable actuators for soft robots:
  Structures, materials, and functionalities},
\newblock \bibinfo{journal}{Advanced Materials} \bibinfo{volume}{34}
  (\bibinfo{year}{2022}) \bibinfo{pages}{2110384}.
\bibitem[{Cao et~al.(2021)Cao, Derakhshani, Fang, Huang, and
  Cao}]{cao2021bistable}
\bibinfo{author}{Y.~Cao}, \bibinfo{author}{M.~Derakhshani},
  \bibinfo{author}{Y.~Fang}, \bibinfo{author}{G.~Huang},
  \bibinfo{author}{C.~Cao},
\newblock \bibinfo{title}{Bistable structures for advanced functional systems},
\newblock \bibinfo{journal}{Advanced Functional Materials} \bibinfo{volume}{31}
  (\bibinfo{year}{2021}) \bibinfo{pages}{2106231}.
\bibitem[{Zhang et~al.(2024)Zhang, Yin, Chen, Ju, Hao, Wu, Sun, xiao Yang, and
  Xu}]{zhang2024review}
\bibinfo{author}{C.~Zhang}, \bibinfo{author}{X.~Yin},
  \bibinfo{author}{R.~Chen}, \bibinfo{author}{K.~Ju}, \bibinfo{author}{Y.~Hao},
  \bibinfo{author}{T.~Wu}, \bibinfo{author}{J.~Sun}, \bibinfo{author}{H.~xiao
  Yang}, \bibinfo{author}{Y.~Xu},
\newblock \bibinfo{title}{A review on reprogrammable bistable structures},
\newblock \bibinfo{journal}{Smart Materials and Structures}
  \bibinfo{volume}{33} (\bibinfo{year}{2024}) \bibinfo{pages}{093001}.
\bibitem[{Scarselli et~al.(2016)Scarselli, Nicassio, Pinto, Ciampa, Iervolino,
  and Meo}]{scarselli2016novel}
\bibinfo{author}{G.~Scarselli}, \bibinfo{author}{F.~Nicassio},
  \bibinfo{author}{F.~Pinto}, \bibinfo{author}{F.~Ciampa},
  \bibinfo{author}{O.~Iervolino}, \bibinfo{author}{M.~Meo},
\newblock \bibinfo{title}{A novel bistable energy harvesting concept},
\newblock \bibinfo{journal}{Smart Materials and Structures}
  \bibinfo{volume}{25} (\bibinfo{year}{2016}) \bibinfo{pages}{055001}.
\bibitem[{Pellegrini et~al.(2013)Pellegrini, Tolou, Schenk, and
  Herder}]{pellegrini2013bistable}
\bibinfo{author}{S.~P. Pellegrini}, \bibinfo{author}{N.~Tolou},
  \bibinfo{author}{M.~Schenk}, \bibinfo{author}{J.~L. Herder},
\newblock \bibinfo{title}{Bistable vibration energy harvesters: a review},
\newblock \bibinfo{journal}{Journal of Intelligent Material Systems and
  Structures} \bibinfo{volume}{24} (\bibinfo{year}{2013})
  \bibinfo{pages}{1303--1312}.
\bibitem[{Kochmann and Bertoldi(2017)}]{kochmann2017exploiting}
\bibinfo{author}{D.~M. Kochmann}, \bibinfo{author}{K.~Bertoldi},
\newblock \bibinfo{title}{Exploiting microstructural instabilities in solids
  and structures: from metamaterials to structural transitions},
\newblock \bibinfo{journal}{Applied mechanics reviews} \bibinfo{volume}{69}
  (\bibinfo{year}{2017}) \bibinfo{pages}{050801}.
\bibitem[{Hussein et~al.(2014)Hussein, Leamy, and
  Ruzzene}]{hussein2014dynamics}
\bibinfo{author}{M.~I. Hussein}, \bibinfo{author}{M.~J. Leamy},
  \bibinfo{author}{M.~Ruzzene},
\newblock \bibinfo{title}{Dynamics of phononic materials and structures:
  Historical origins, recent progress, and future outlook},
\newblock \bibinfo{journal}{Applied Mechanics Reviews} \bibinfo{volume}{66}
  (\bibinfo{year}{2014}) \bibinfo{pages}{040802}.
\bibitem[{Khelif et~al.(2004)Khelif, Choujaa, Benchabane, Djafari-Rouhani, and
  Laude}]{khelif2004guiding}
\bibinfo{author}{A.~Khelif}, \bibinfo{author}{A.~Choujaa},
  \bibinfo{author}{S.~Benchabane}, \bibinfo{author}{B.~Djafari-Rouhani},
  \bibinfo{author}{V.~Laude},
\newblock \bibinfo{title}{Guiding and bending of acoustic waves in highly
  confined phononic crystal waveguides},
\newblock \bibinfo{journal}{Applied physics letters} \bibinfo{volume}{84}
  (\bibinfo{year}{2004}) \bibinfo{pages}{4400--4402}.
\bibitem[{Kafesaki et~al.(2000)Kafesaki, Sigalas, and
  Garcia}]{kafesaki2000frequency}
\bibinfo{author}{M.~Kafesaki}, \bibinfo{author}{M.~Sigalas},
  \bibinfo{author}{N.~Garcia},
\newblock \bibinfo{title}{Frequency modulation in the transmittivity of wave
  guides in elastic-wave band-gap materials},
\newblock \bibinfo{journal}{Physical Review Letters} \bibinfo{volume}{85}
  (\bibinfo{year}{2000}) \bibinfo{pages}{4044}.
\bibitem[{Cummer and Schurig(2007)}]{cummer2007one}
\bibinfo{author}{S.~A. Cummer}, \bibinfo{author}{D.~Schurig},
\newblock \bibinfo{title}{One path to acoustic cloaking},
\newblock \bibinfo{journal}{New journal of physics} \bibinfo{volume}{9}
  (\bibinfo{year}{2007}) \bibinfo{pages}{45}.
\bibitem[{Elser et~al.(2006)Elser, Andersen, Korn, Gl{\"o}ckl, Lorenz,
  Marquardt, and Leuchs}]{elser2006reduction}
\bibinfo{author}{D.~Elser}, \bibinfo{author}{U.~Andersen},
  \bibinfo{author}{A.~Korn}, \bibinfo{author}{O.~Gl{\"o}ckl},
  \bibinfo{author}{S.~Lorenz}, \bibinfo{author}{C.~Marquardt},
  \bibinfo{author}{G.~Leuchs},
\newblock \bibinfo{title}{Reduction of guided acoustic wave brillouin
  scattering in photonic crystal fibers},
\newblock \bibinfo{journal}{Physical review letters} \bibinfo{volume}{97}
  (\bibinfo{year}{2006}) \bibinfo{pages}{133901}.
\bibitem[{Casadei et~al.(2010)Casadei, Dozio, Ruzzene, and
  Cunefare}]{casadei2010periodic}
\bibinfo{author}{F.~Casadei}, \bibinfo{author}{L.~Dozio},
  \bibinfo{author}{M.~Ruzzene}, \bibinfo{author}{K.~A. Cunefare},
\newblock \bibinfo{title}{Periodic shunted arrays for the control of noise
  radiation in an enclosure},
\newblock \bibinfo{journal}{Journal of sound and vibration}
  \bibinfo{volume}{329} (\bibinfo{year}{2010}) \bibinfo{pages}{3632--3646}.
\bibitem[{Betts et~al.(2013)Betts, Bowen, Kim, Gathercole, Clarke, and
  Inman}]{Betts2013}
\bibinfo{author}{D.~Betts}, \bibinfo{author}{C.~Bowen},
  \bibinfo{author}{H.~Kim}, \bibinfo{author}{N.~Gathercole},
  \bibinfo{author}{C.~Clarke}, \bibinfo{author}{D.~Inman},
\newblock \bibinfo{title}{Nonlinear dynamics of a bistable
  piezoelectric-composite energy harvester for broadband application},
\newblock \bibinfo{journal}{European Physical Journal Special Topics}
  \bibinfo{volume}{222} (\bibinfo{year}{2013}) \bibinfo{pages}{1553--1562}.
\bibitem[{Yang et~al.(2014)Yang, Harne, Wang, and Huang}]{Yang2014}
\bibinfo{author}{K.~Yang}, \bibinfo{author}{R.~L. Harne},
  \bibinfo{author}{K.~W. Wang}, \bibinfo{author}{H.~Huang},
\newblock \bibinfo{title}{Dynamic stabilization of a bistable suspension system
  attached to a flexible host structure for operational safety enhancement},
\newblock \bibinfo{journal}{Journal of Sound and Vibration}
  \bibinfo{volume}{333} (\bibinfo{year}{2014}) \bibinfo{pages}{6651--6661}.
\bibitem[{Harne et~al.(2013)Harne, Thota, and Wang}]{Harne2013}
\bibinfo{author}{R.~L. Harne}, \bibinfo{author}{M.~Thota},
  \bibinfo{author}{K.~W. Wang},
\newblock \bibinfo{title}{Concise and high-fidelity predictive criteria for
  maximizing performance and robustness of bistable energy harvesters},
\newblock \bibinfo{journal}{Applied Physics Letters} \bibinfo{volume}{102}
  (\bibinfo{year}{2013}) \bibinfo{pages}{053903}.
\bibitem[{Wu et~al.(2014)Wu, Harne, and Wang}]{Wu2014}
\bibinfo{author}{Z.~Z. Wu}, \bibinfo{author}{R.~L. Harne},
  \bibinfo{author}{K.~W. Wang},
\newblock \bibinfo{title}{Energy harvester synthesis via coupled
  linear-bistable system with multistable dynamics},
\newblock \bibinfo{journal}{ASME Journal of Applied Mechanics}
  \bibinfo{volume}{81} (\bibinfo{year}{2014}) \bibinfo{pages}{061005}.
\bibitem[{Johnson et~al.(2014)Johnson, Harne, and Wang}]{Johnson2014}
\bibinfo{author}{D.~R. Johnson}, \bibinfo{author}{R.~L. Harne},
  \bibinfo{author}{K.~W. Wang},
\newblock \bibinfo{title}{A disturbance cancellation perspective on vibration
  control using a bistable snap-through attachment},
\newblock \bibinfo{journal}{ASME Journal of Vibration and Acoustics}
  \bibinfo{volume}{136} (\bibinfo{year}{2014}) \bibinfo{pages}{031006}.
\bibitem[{Nadkarni et~al.(2016{\natexlab{a}})Nadkarni, Daraio, Abeyaratne, and
  Kochmann}]{Nadkarni2016a}
\bibinfo{author}{N.~Nadkarni}, \bibinfo{author}{C.~Daraio},
  \bibinfo{author}{R.~Abeyaratne}, \bibinfo{author}{D.~M. Kochmann},
\newblock \bibinfo{title}{Universal energy transport law for dissipative and
  diffusive phase transitions},
\newblock \bibinfo{journal}{Physical Review B} \bibinfo{volume}{93}
  (\bibinfo{year}{2016}{\natexlab{a}}) \bibinfo{pages}{104109}.
\bibitem[{Nadkarni et~al.(2016{\natexlab{b}})Nadkarni, Arrieta, Chong,
  Kochmann, and Daraio}]{Nadkarni2016b}
\bibinfo{author}{N.~Nadkarni}, \bibinfo{author}{A.~F. Arrieta},
  \bibinfo{author}{C.~Chong}, \bibinfo{author}{D.~M. Kochmann},
  \bibinfo{author}{C.~Daraio},
\newblock \bibinfo{title}{Unidirectional transition waves in bistable
  lattices},
\newblock \bibinfo{journal}{Physical Review Letters} \bibinfo{volume}{116}
  (\bibinfo{year}{2016}{\natexlab{b}}) \bibinfo{pages}{244501}.
\bibitem[{Shim et~al.(2012)Shim, Perdigou, Chen, Bertoldi, and
  Reis}]{shim2012buckling}
\bibinfo{author}{J.~Shim}, \bibinfo{author}{C.~Perdigou},
  \bibinfo{author}{E.~R. Chen}, \bibinfo{author}{K.~Bertoldi},
  \bibinfo{author}{P.~M. Reis},
\newblock \bibinfo{title}{Buckling-induced encapsulation of structured elastic
  shells under pressure},
\newblock \bibinfo{journal}{Proceedings of the National Academy of Sciences}
  \bibinfo{volume}{109} (\bibinfo{year}{2012}) \bibinfo{pages}{5978--5983}.
\bibitem[{Chen et~al.(2021)Chen, Panetta, Schnaubelt, and
  Pauly}]{chen2021bistable}
\bibinfo{author}{T.~Chen}, \bibinfo{author}{J.~Panetta},
  \bibinfo{author}{M.~Schnaubelt}, \bibinfo{author}{M.~Pauly},
\newblock \bibinfo{title}{Bistable auxetic surface structures},
\newblock \bibinfo{journal}{ACM Transactions on Graphics (TOG)}
  \bibinfo{volume}{40} (\bibinfo{year}{2021}) \bibinfo{pages}{1--9}.
\bibitem[{Grima and Gatt(2010)}]{grima2010perforated}
\bibinfo{author}{J.~N. Grima}, \bibinfo{author}{R.~Gatt},
\newblock \bibinfo{title}{Perforated sheets exhibiting negative poisson's
  ratios},
\newblock \bibinfo{journal}{Advanced engineering materials}
  \bibinfo{volume}{12} (\bibinfo{year}{2010}) \bibinfo{pages}{460--464}.
\bibitem[{Shan et~al.(2015)Shan, Kang, Zhao, Fang, and
  Bertoldi}]{shan2015design}
\bibinfo{author}{S.~Shan}, \bibinfo{author}{S.~H. Kang},
  \bibinfo{author}{Z.~Zhao}, \bibinfo{author}{L.~Fang},
  \bibinfo{author}{K.~Bertoldi},
\newblock \bibinfo{title}{Design of planar isotropic negative poisson’s ratio
  structures},
\newblock \bibinfo{journal}{Extreme Mechanics Letters} \bibinfo{volume}{4}
  (\bibinfo{year}{2015}) \bibinfo{pages}{96--102}.
\bibitem[{Grima et~al.(2016)Grima, Mizzi, Azzopardi, Gatt
  et~al.}]{grima2016auxetic}
\bibinfo{author}{J.~N. Grima}, \bibinfo{author}{L.~Mizzi},
  \bibinfo{author}{K.~M. Azzopardi}, \bibinfo{author}{R.~Gatt}, et~al.,
\newblock \bibinfo{title}{Auxetic perforated mechanical metamaterials with
  randomly oriented cuts},
\newblock \bibinfo{journal}{Adv. Mater} \bibinfo{volume}{28}
  (\bibinfo{year}{2016}) \bibinfo{pages}{385--389}.
\bibitem[{Rafsanjani and Pasini(2016)}]{rafsanjani2016bistable}
\bibinfo{author}{A.~Rafsanjani}, \bibinfo{author}{D.~Pasini},
\newblock \bibinfo{title}{Bistable auxetic mechanical metamaterials inspired by
  ancient geometric motifs},
\newblock \bibinfo{journal}{Extreme Mechanics Letters} \bibinfo{volume}{9}
  (\bibinfo{year}{2016}) \bibinfo{pages}{291--296}.
\bibitem[{Deng et~al.(2020)Deng, Wang, Tournat, and Bertoldi}]{Deng2020}
\bibinfo{author}{B.~Deng}, \bibinfo{author}{P.~Wang},
  \bibinfo{author}{V.~Tournat}, \bibinfo{author}{K.~Bertoldi},
\newblock \bibinfo{title}{{Nonlinear Transition Waves in Free-Standing Bistable
  Chains}, journal = {Journal of the Mechanics and Physics of Solids}}
  \bibinfo{volume}{137} (\bibinfo{year}{2020}) \bibinfo{pages}{103826}.
  \DOIprefix\doi{10.1016/j.jmps.2019.103826}.
\bibitem[{Yasuda et~al.(2020)Yasuda, Raney, Tournat, and Bertoldi}]{Yasuda2020}
\bibinfo{author}{H.~Yasuda}, \bibinfo{author}{J.~R. Raney},
  \bibinfo{author}{V.~Tournat}, \bibinfo{author}{K.~Bertoldi},
\newblock \bibinfo{title}{{Transition Waves and Formation of Domain Walls in
  Multistable Mechanical Metamaterials}},
\newblock \bibinfo{journal}{Physical Review Applied} \bibinfo{volume}{13}
  (\bibinfo{year}{2020}) \bibinfo{pages}{054067}.
  \DOIprefix\doi{10.1103/PhysRevApplied.13.054067}.
\bibitem[{Jin et~al.(2020)Jin, Khajehtourian, Mueller, Rafsanjani, Tournat,
  Bertoldi, and Kochmann}]{jin2020guided}
\bibinfo{author}{L.~Jin}, \bibinfo{author}{R.~Khajehtourian},
  \bibinfo{author}{J.~Mueller}, \bibinfo{author}{A.~Rafsanjani},
  \bibinfo{author}{V.~Tournat}, \bibinfo{author}{K.~Bertoldi},
  \bibinfo{author}{D.~M. Kochmann},
\newblock \bibinfo{title}{Guided transition waves in multistable mechanical
  metamaterials},
\newblock \bibinfo{journal}{Proceedings of the National Academy of Sciences}
  \bibinfo{volume}{117} (\bibinfo{year}{2020}) \bibinfo{pages}{2319--2325}.
\bibitem[{Bonthron and Tubaldi(2025)}]{Bonthron2025}
\bibinfo{author}{M.~Bonthron}, \bibinfo{author}{E.~Tubaldi},
\newblock \bibinfo{title}{{Transition Wave Interference in Multistable
  Mechanical Metamaterials}},
\newblock \bibinfo{journal}{Journal of Applied Mechanics} \bibinfo{volume}{92}
  (\bibinfo{year}{2025}) \bibinfo{pages}{051005}.
  \DOIprefix\doi{10.1115/1.4065775}.
\bibitem[{Paliovaios et~al.(2024)Paliovaios, Theocharis, and
  Achilleos}]{Paliovaios2024}
\bibinfo{author}{A.~Paliovaios}, \bibinfo{author}{G.~Theocharis},
  \bibinfo{author}{V.~Achilleos},
\newblock \bibinfo{title}{{Transition Waves in Bistable Systems Generated by
  Collision of Moving Breathers}},
\newblock \bibinfo{journal}{Extreme Mechanics Letters} \bibinfo{volume}{66}
  (\bibinfo{year}{2024}) \bibinfo{pages}{102020}.
  \DOIprefix\doi{10.1016/j.eml.2024.102020}.
\bibitem[{Frazier and Kochmann(2023)}]{Frazier2023}
\bibinfo{author}{M.~J. Frazier}, \bibinfo{author}{D.~M. Kochmann},
\newblock \bibinfo{title}{{Phase Transitions in Hierarchical, Multistable
  Mechanical Metamaterials}},
\newblock \bibinfo{journal}{Extreme Mechanics Letters} \bibinfo{volume}{61}
  (\bibinfo{year}{2023}) \bibinfo{pages}{102055}.
  \DOIprefix\doi{10.1016/j.eml.2023.102055}.
\bibitem[{Zhang et~al.(2022)Zhang, Fang, Sun, and Liu}]{Zhang2022}
\bibinfo{author}{Y.~Zhang}, \bibinfo{author}{X.~Fang},
  \bibinfo{author}{J.~Sun}, \bibinfo{author}{Z.~Liu},
\newblock \bibinfo{title}{{Modeling and Analysis of Transition Waves in
  Bistable Mechanical Chains}},
\newblock \bibinfo{journal}{International Journal of Solids and Structures}
  \bibinfo{volume}{251} (\bibinfo{year}{2022}) \bibinfo{pages}{111667}.
  \DOIprefix\doi{10.1016/j.ijsolstr.2022.111667}.
\bibitem[{Friedrich et~al.(2018)Friedrich, Pfeiffer, and
  Gengnagel}]{friedrich2018locally}
\bibinfo{author}{J.~Friedrich}, \bibinfo{author}{S.~Pfeiffer},
  \bibinfo{author}{C.~Gengnagel},
\newblock \bibinfo{title}{Locally varied auxetic structures for doubly-curved
  shapes},
\newblock in: \bibinfo{booktitle}{Humanizing Digital Reality: Design Modelling
  Symposium Paris 2017}, \bibinfo{organization}{Springer},
  \bibinfo{year}{2018}, pp. \bibinfo{pages}{323--336}.
\bibitem[{Konakovi{\'c} et~al.(2016)Konakovi{\'c}, Crane, Deng, Bouaziz, Piker,
  and Pauly}]{konakovic2016beyond}
\bibinfo{author}{M.~Konakovi{\'c}}, \bibinfo{author}{K.~Crane},
  \bibinfo{author}{B.~Deng}, \bibinfo{author}{S.~Bouaziz},
  \bibinfo{author}{D.~Piker}, \bibinfo{author}{M.~Pauly},
\newblock \bibinfo{title}{Beyond developable: computational design and
  fabrication with auxetic materials},
\newblock \bibinfo{journal}{ACM Transactions On Graphics (TOG)}
  \bibinfo{volume}{35} (\bibinfo{year}{2016}) \bibinfo{pages}{1--11}.
\bibitem[{Konakovi{\'c}-Lukovi{\'c} et~al.(2018)Konakovi{\'c}-Lukovi{\'c},
  Panetta, Crane, and Pauly}]{konakovic2018rapid}
\bibinfo{author}{M.~Konakovi{\'c}-Lukovi{\'c}}, \bibinfo{author}{J.~Panetta},
  \bibinfo{author}{K.~Crane}, \bibinfo{author}{M.~Pauly},
\newblock \bibinfo{title}{Rapid deployment of curved surfaces via programmable
  auxetics},
\newblock \bibinfo{journal}{ACM Transactions on Graphics (TOG)}
  \bibinfo{volume}{37} (\bibinfo{year}{2018}) \bibinfo{pages}{1--13}.
\bibitem[{Khajehtourian and Kochmann(2021)}]{khajehtourian2021continuum}
\bibinfo{author}{R.~Khajehtourian}, \bibinfo{author}{D.~M. Kochmann},
\newblock \bibinfo{title}{A continuum description of substrate-free dissipative
  reconfigurable metamaterials},
\newblock \bibinfo{journal}{Journal of the Mechanics and Physics of Solids}
  \bibinfo{volume}{147} (\bibinfo{year}{2021}) \bibinfo{pages}{104217}.
\bibitem[{Rivlin(1948)}]{rivlin1948large}
\bibinfo{author}{R.~S. Rivlin},
\newblock \bibinfo{title}{Large elastic deformations of isotropic materials iv.
  further developments of the general theory},
\newblock \bibinfo{journal}{Philosophical transactions of the royal society of
  London. Series A, Mathematical and physical sciences} \bibinfo{volume}{241}
  (\bibinfo{year}{1948}) \bibinfo{pages}{379--397}.
\bibitem[{Ogden(1972)}]{ogden1972large}
\bibinfo{author}{R.~W. Ogden},
\newblock \bibinfo{title}{Large deformation isotropic elasticity--on the
  correlation of theory and experiment for incompressible rubberlike solids},
\newblock \bibinfo{journal}{Proceedings of the Royal Society of London. A.
  Mathematical and Physical Sciences} \bibinfo{volume}{326}
  (\bibinfo{year}{1972}) \bibinfo{pages}{565--584}.
\bibitem[{Ogden(1997)}]{ogden1997non}
\bibinfo{author}{R.~W. Ogden}, \bibinfo{title}{Non-linear elastic
  deformations}, \bibinfo{publisher}{Courier Corporation},
  \bibinfo{year}{1997}.
\bibitem[{Ogden(1986)}]{ogden1986recent}
\bibinfo{author}{R.~W. Ogden},
\newblock \bibinfo{title}{Recent advances in the phenomenological theory of
  rubber elasticity},
\newblock \bibinfo{journal}{Rubber Chemistry and Technology}
  \bibinfo{volume}{59} (\bibinfo{year}{1986}) \bibinfo{pages}{361--383}.
\bibitem[{Criscione et~al.(2000)Criscione, Humphrey, Douglas, and
  Hunter}]{criscione2000invariant}
\bibinfo{author}{J.~C. Criscione}, \bibinfo{author}{J.~D. Humphrey},
  \bibinfo{author}{A.~S. Douglas}, \bibinfo{author}{W.~C. Hunter},
\newblock \bibinfo{title}{An invariant basis for natural strain which yields
  orthogonal stress response terms in isotropic hyperelasticity},
\newblock \bibinfo{journal}{Journal of the Mechanics and Physics of Solids}
  \bibinfo{volume}{48} (\bibinfo{year}{2000}) \bibinfo{pages}{2445--2465}.
\bibitem[{Zauderer(2011)}]{zauderer2011partial}
\bibinfo{author}{E.~Zauderer}, \bibinfo{title}{Partial differential equations
  of applied mathematics}, \bibinfo{publisher}{John Wiley \& Sons},
  \bibinfo{year}{2011}.
\bibitem[{Bazant et~al.(1984)Bazant, Belytschko, Chang
  et~al.}]{bazant1984continuum}
\bibinfo{author}{Z.~P. Bazant}, \bibinfo{author}{T.~B. Belytschko},
  \bibinfo{author}{T.-P. Chang}, et~al.,
\newblock \bibinfo{title}{Continuum theory for strain-softening},
\newblock \bibinfo{journal}{Journal of Engineering Mechanics}
  \bibinfo{volume}{110} (\bibinfo{year}{1984}) \bibinfo{pages}{1666--1692}.
\bibitem[{Aifantis(1984)}]{10.1115/1.3225725}
\bibinfo{author}{E.~C. Aifantis},
\newblock \bibinfo{title}{On the microstructural origin of certain inelastic
  models},
\newblock \bibinfo{journal}{Journal of Engineering Materials and Technology}
  \bibinfo{volume}{106} (\bibinfo{year}{1984}) \bibinfo{pages}{326--330}.
\bibitem[{Peerlings et~al.(1996)Peerlings, De~Borst, Brekelmans, De~Vree, and
  Spee}]{peerlings1996some}
\bibinfo{author}{R.~d. Peerlings}, \bibinfo{author}{R.~De~Borst},
  \bibinfo{author}{W.~d. Brekelmans}, \bibinfo{author}{J.~De~Vree},
  \bibinfo{author}{I.~Spee},
\newblock \bibinfo{title}{Some observations on localisation in non-local and
  gradient damage models},
\newblock \bibinfo{journal}{European Journal of Mechanics. A, Solids}
  \bibinfo{volume}{15} (\bibinfo{year}{1996}) \bibinfo{pages}{937--953}.
\bibitem[{Fleck and Hutchinson(1993)}]{fleck1993phenomenological}
\bibinfo{author}{N.~Fleck}, \bibinfo{author}{J.~Hutchinson},
\newblock \bibinfo{title}{A phenomenological theory for strain gradient effects
  in plasticity},
\newblock \bibinfo{journal}{Journal of the Mechanics and Physics of Solids}
  \bibinfo{volume}{41} (\bibinfo{year}{1993}) \bibinfo{pages}{1825--1857}.
\bibitem[{Gurtin(2003)}]{gurtin2003framework}
\bibinfo{author}{M.~E. Gurtin},
\newblock \bibinfo{title}{On a framework for small-deformation viscoplasticity:
  free energy, microforces, strain gradients},
\newblock \bibinfo{journal}{International Journal of Plasticity}
  \bibinfo{volume}{19} (\bibinfo{year}{2003}) \bibinfo{pages}{47--90}.
\bibitem[{Forest(2009)}]{forest2009micromorphic}
\bibinfo{author}{S.~Forest},
\newblock \bibinfo{title}{Micromorphic approach for gradient elasticity,
  viscoplasticity, and damage},
\newblock \bibinfo{journal}{Journal of Engineering Mechanics}
  \bibinfo{volume}{135} (\bibinfo{year}{2009}) \bibinfo{pages}{117--131}.
\bibitem[{Shaat et~al.(2020)Shaat, Ghavanloo, and Fazelzadeh}]{shaat2020review}
\bibinfo{author}{M.~Shaat}, \bibinfo{author}{E.~Ghavanloo},
  \bibinfo{author}{S.~A. Fazelzadeh},
\newblock \bibinfo{title}{Review on nonlocal continuum mechanics: physics,
  material applicability, and mathematics},
\newblock \bibinfo{journal}{Mechanics of Materials} \bibinfo{volume}{150}
  (\bibinfo{year}{2020}) \bibinfo{pages}{103587}.
\bibitem[{Needleman(1988)}]{needleman1988material}
\bibinfo{author}{A.~Needleman},
\newblock \bibinfo{title}{Material rate dependence and mesh sensitivity in
  localization problems},
\newblock \bibinfo{journal}{Computer Methods in Applied Mechanics and
  Engineering} \bibinfo{volume}{67} (\bibinfo{year}{1988})
  \bibinfo{pages}{69--85}.
\bibitem[{Loret and Prevost(1990)}]{loret1990dynamic}
\bibinfo{author}{B.~Loret}, \bibinfo{author}{J.~H. Prevost},
\newblock \bibinfo{title}{{Dynamic strain localization in elasto-(visco-)
  plastic solids. Part 1: General formulation and one-dimensional examples}},
\newblock \bibinfo{journal}{Computer Methods in Applied Mechanics and
  Engineering} \bibinfo{volume}{83} (\bibinfo{year}{1990})
  \bibinfo{pages}{247--273}.
\bibitem[{Prevost and Loret(1990)}]{prevost1990dynamic}
\bibinfo{author}{J.~H. Prevost}, \bibinfo{author}{B.~Loret},
\newblock \bibinfo{title}{{Dynamic strain localization in elasto-(visco-)
  plastic solids. Part 2: plane strain examples}},
\newblock \bibinfo{journal}{Computer Methods in Applied Mechanics and
  Engineering} \bibinfo{volume}{83} (\bibinfo{year}{1990})
  \bibinfo{pages}{275--294}.
\bibitem[{Wang et~al.(1996)Wang, Sluys, and De~Borst}]{wang1996interaction}
\bibinfo{author}{W.~Wang}, \bibinfo{author}{L.~J. Sluys},
  \bibinfo{author}{R.~De~Borst},
\newblock \bibinfo{title}{Interaction between material length scale and
  imperfection size for localisation phenomena in viscoplastic media},
\newblock \bibinfo{journal}{European Journal of Mechanics. A, Solids}
  \bibinfo{volume}{15} (\bibinfo{year}{1996}) \bibinfo{pages}{447--464}.
\bibitem[{Benallal(2008)}]{benallal2008note}
\bibinfo{author}{A.~Benallal},
\newblock \bibinfo{title}{A note on ill-posedness for rate-dependent problems
  and its relation to the rate-independent case},
\newblock \bibinfo{journal}{Computational Mechanics} \bibinfo{volume}{42}
  (\bibinfo{year}{2008}) \bibinfo{pages}{261--269}.
\bibitem[{Belytschko et~al.(1991)Belytschko, Moran, and
  Kulkarni}]{10.1115/1.2897246}
\bibinfo{author}{T.~Belytschko}, \bibinfo{author}{B.~Moran},
  \bibinfo{author}{M.~Kulkarni},
\newblock \bibinfo{title}{On the crucial role of imperfections in quasi-static
  viscoplastic solutions},
\newblock \bibinfo{journal}{Journal of Applied Mechanics} \bibinfo{volume}{58}
  (\bibinfo{year}{1991}) \bibinfo{pages}{658--665}.
\bibitem[{Aravas and Papadioti(2021)}]{aravas2021non}
\bibinfo{author}{N.~Aravas}, \bibinfo{author}{I.~Papadioti},
\newblock \bibinfo{title}{{A non-local plasticity model for porous metals with
  deformation-induced anisotropy: Mathematical and computational issues}},
\newblock \bibinfo{journal}{Journal of the Mechanics and Physics of Solids}
  \bibinfo{volume}{146} (\bibinfo{year}{2021}) \bibinfo{pages}{104190}.
\bibitem[{Aravas and Xenos(2023)}]{aravas2023implicit}
\bibinfo{author}{N.~Aravas}, \bibinfo{author}{S.~Xenos},
\newblock \bibinfo{title}{{“Implicit” vs “Explicit” gradient plasticity
  models: Do they always remove mesh dependence in softening materials?}},
\newblock \bibinfo{journal}{International Journal of Solids and Structures}
  \bibinfo{volume}{281} (\bibinfo{year}{2023}) \bibinfo{pages}{112415}.
\bibitem[{Danielsson et~al.(2002)Danielsson, Parks, and
  Boyce}]{danielsson2002three}
\bibinfo{author}{M.~Danielsson}, \bibinfo{author}{D.~Parks},
  \bibinfo{author}{M.~C. Boyce},
\newblock \bibinfo{title}{Three-dimensional micromechanical modeling of voided
  polymeric materials},
\newblock \bibinfo{journal}{Journal of the Mechanics and Physics of Solids}
  \bibinfo{volume}{50} (\bibinfo{year}{2002}) \bibinfo{pages}{351--379}.
\bibitem[{Liu et~al.(2022)Liu, Pratapa, Misseroni, Tachi, and
  Paulino}]{liu2022triclinic}
\bibinfo{author}{K.~Liu}, \bibinfo{author}{P.~P. Pratapa},
  \bibinfo{author}{D.~Misseroni}, \bibinfo{author}{T.~Tachi},
  \bibinfo{author}{G.~H. Paulino},
\newblock \bibinfo{title}{Triclinic metamaterials by tristable origami with
  reprogrammable frustration},
\newblock \bibinfo{journal}{Advanced Materials} \bibinfo{volume}{34}
  (\bibinfo{year}{2022}) \bibinfo{pages}{2107998}.
\bibitem[{Hutchinson and Neale(1977)}]{hutchinson1977influence}
\bibinfo{author}{J.~Hutchinson}, \bibinfo{author}{K.~Neale},
\newblock \bibinfo{title}{Influence of strain-rate sensitivity on necking under
  uniaxial tension},
\newblock \bibinfo{journal}{Acta Metallurgica} \bibinfo{volume}{25}
  (\bibinfo{year}{1977}) \bibinfo{pages}{839--846}.
\bibitem[{Tvergaard and Needleman(1980)}]{10.1115/1.3153742}
\bibinfo{author}{V.~Tvergaard}, \bibinfo{author}{A.~Needleman},
\newblock \bibinfo{title}{On the localization of buckling patterns},
\newblock \bibinfo{journal}{Journal of Applied Mechanics} \bibinfo{volume}{47}
  (\bibinfo{year}{1980}) \bibinfo{pages}{613--619}.
\bibitem[{Kelly and Lai(2011)}]{kelly2011use}
\bibinfo{author}{J.~Kelly}, \bibinfo{author}{J.-W. Lai},
\newblock \bibinfo{title}{The use of tests on high-shape-factor bearings to
  estimate the bulk modulus of natural rubber},
\newblock \bibinfo{journal}{Seismic Isolation and Protection Systems}
  \bibinfo{volume}{2} (\bibinfo{year}{2011}) \bibinfo{pages}{21--33}.

\end{thebibliography}
\end{document}